\documentclass[twocolumn,showpacs,preprintnumbers,amsmath,amssymb,prb]{revtex4}

\usepackage{graphicx}
\usepackage{dcolumn}
\usepackage{longtable}
\usepackage{bm}
\usepackage{epsfig}

\begin{document}

\title{X-ray Linear Dichroism in cubic compounds: the case of Cr$^{3+}$ in MgAl$_{2}$O$_{4}$} 
\author{{Am\'{e}lie Juhin}$^{1}$} 
\email{amelie.juhin@impmc.jussieu.fr} 
\author{{Christian Brouder}$^{1}$} 
\author{{Marie-Anne Arrio}$^{1}$} 
\author{{Delphine Cabaret}$^{1}$} 
\author{{Philippe Sainctavit}$^{1}$} 
\author{{Etienne Balan}$^{1,2}$}
\author{{Am\'{e}lie Bordage}$^{1}$} 
\author{{Georges Calas}$^{1}$} 
\author{{Sigrid G. Eeckhout}$^{3}$} 
\author{{Pieter Glatzel}$^{3}$}

\affiliation{$^{1}$ Institut de Min\'eralogie et 
 de Physique des Milieux Condens\'es (IMPMC), UMR CNRS 7590,
 Universit\'e Pierre et Marie Curie, Paris 6 $\&$ Paris 7, IPGP,
4 place Jussieu, 75052 Paris Cedex 05, France\\ 
$^{2}$ Institut de Recherche pour le D\'eveloppement (IRD), UR T058, 
  213 rue La Fayette, 75480 Paris Cedex 10, France\\
 $^{3}$ European Synchrotron Radiation Facility, 6 rue Jules Horowitz, BP 220, 38043 Grenoble Cedex, France}

\begin{abstract}
The angular dependence (x-ray linear dichroism) of the Cr K pre-edge in 
MgAl$_{2}$O$_{4}$:Cr$^{3+}$ spinel is measured by
means of x-ray absorption near edge structure spectroscopy (XANES) and compared to
calculations based on density functional theory (DFT) and ligand field multiplet theory (LFM).
We also present an efficient method, based on symmetry considerations, to compute the dichroism
of the cubic crystal starting from the dichroism of a single substitutional site.
DFT shows that the electric dipole transitions do not contribute to the features visible in the pre-edge and  provides a clear vision of the assignment of the 1\textit{s}$\rightarrow$3\textit{d} transitions.
However, DFT is unable to reproduce quantitatively the angular dependence of the pre-edge, which is, on the other side, well reproduced by LFM calculations.
The most relevant factors determining the dichroism of Cr K pre-edge are identified as the site distortion and 3\textit{d}-3\textit{d} electronic repulsion.
From this combined DFT, LFM approach is concluded that
when the pre-edge features are more intense than 4~\% of the edge jump,
pure quadrupole transitions cannot explain alone the origin of the pre-edge.
Finally, the shape of the dichroic signal is more sensitive than the isotropic spectrum
to the trigonal distortion of the substitutional site.
This suggests the possibility to obtain quantitative information on site distortion from the x-ray linear dichroism by performing angular dependent measurements on single crystals.
\end{abstract} 


\maketitle

\section{Introduction}
Transition metal elements play an essential role in physics (magnetic materials, superconductors...), coordination chemistry (catalysis, metalloproteins) or geophysics (3\textit{d} elements are major constituents of the Earth and planets). To understand the properties that transition elements impart to the materials they are inserted in, X-ray Absorption Near Edge Structure (XANES) spectroscopy has been widely used, since it provides unique information on their local surrounding and electronic structure. In particular, the position and intensity of the localized transitions observed at the K pre-edge (1\textit{s}$\rightarrow$3\textit{d} transitions) are sensitive to the cation oxidation state, the geometry of its environment (coordination number and symmetry), and the degree of admixture between \textit{p} and \textit{d} orbitals. For example, the shape and area of the pre-edge are commonly used to quantify the redox states of transition elements in crystals, glasses and coordination complexes, by comparison to those recorded on reference compounds.\cite{Balan,Eeckhout,Galoisy} However, it is not straightforward to obtain this kind of information on single crystals. Indeed, it is well known that the XANES spectra of non-cubic crystals show an angular dependence, when the polarization and the direction of the incident x-ray beam (here, designated as unit vectors, $\bm{\mathrm{\hat{\varepsilon}}}$ and $\bm{\mathrm{\hat{k}}}$, respectively) are varied. For cubic crystals, the problem may seem at first sight more simple. Electric dipole transitions (e.g., 1\textit{s}$\rightarrow$\textit{p} transitions) are isotropic. They contribute mainly to the edge, but also to the pre-edge if one of the three following situations is encountered :  (i) there is \textit{p}-\textit{d} intrasite hybridization (e.g., the crystallographic site does not show an inversion center), (ii) the thermally activated vibrations remove the inversion center, (iii) there is \textit{p}-\textit{d} intersite hybridization (in samples highly concentrated in the investigated element). Electric quadrupole transitions are anisotropic and the cubic crystal thus shows an angular dependence. The information carried by the pre-edge features can be derived for cubic crystals by taking advantage of this angular dependence. In particular, the respective proportion of electric dipole and quadrupole transitions in the pre-edge can be derived, by measuring XANES spectra for various known orientations ($\bm{\mathrm{\hat{\varepsilon}}}$,$\bm{\mathrm{\hat{k}}}$) of the incident beam.\cite{Drager}  In addition, the symmetry of the crystallographic sites, that host the investigated element, is often a subgroup of the cubic group. The number of equivalent sites is given by the ratio of the multiplicity of the space group and the multiplicity of the point group. The XANES spectrum of the cubic crystal is thus the average over the equivalent sites of the individual site spectra. Hence, the derivation of structural and electronic information for a single site is not straightforward, which makes the use of group theory and theoretical computations mandatory. 

\begin{figure}[!t]
\includegraphics[width=7.9cm]{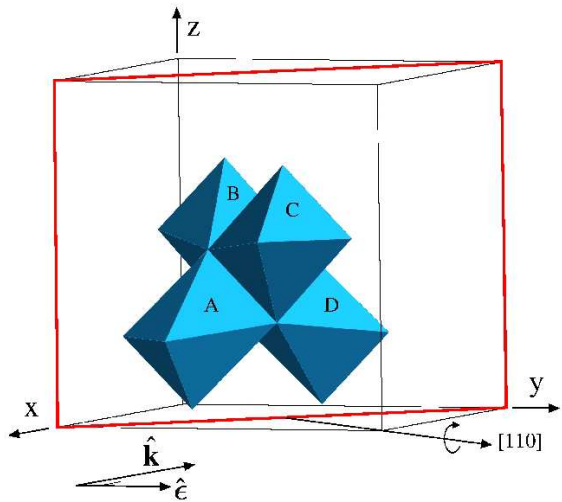}
\caption{\label{fig1} (Color online) Cubic cell of the spinel structure and experimental setup. The four equivalent octahedral sites are labeled according to the coordinates given in Table ~\ref{tab:coord}. Site distortion has been slightly exagerated for clarity. The sample is cut along the (110) plane (red) and  rotated along the [110] direction, while $\bm{\mathrm{\hat{\varepsilon}}}$ and $\bm{\mathrm{\hat{k}}}$ are fixed in the laboratory system. The figure corresponds to the experimental setup taken as starting point ($\alpha_{rot}$ = 0$^{\circ}$). For this configuration, the [001] axis of the cube is in the vertical plane, perpendicular to $\bm{\mathrm{\hat{\varepsilon}}}$~=~[010] and $\bm{\mathrm{\hat{k}}}$~=~[$\bar{1}$00].}
\end{figure}

Among cubic oxides, spinels have attracted considerable interest for their optical, electronic, mechanical and magnetic properties.\cite{Ueda,Rodriguez,Schiessl} In the Earth's interior, the formation of silicate spinels has major geophysical implications.\cite{Burnley} More specifically, MgAl$_{2}$O$_{4}$ spinels are used in a broad range of applications, including optically transparent materials, catalyst supports, nuclear waste management and cement castables.\cite{Guo,Beauvy,Mukho} Cr$^{3+}$ often substitutes for Al$^{3+}$ in MgAl$_{2}$O$_{4}$, which causes a red color and makes natural Cr-spinels valuable gemstones. Cr$^{3+}$ is intentionally added to high-temperature refractory materials to improve their thermal and mechanical properties.\cite{Levy} In MgAl$_{2}$O$_{4}$ spinel ($\mathit{Fd\bar{3}m}$ space group symmetry), Al$^{3+}$ cations occur at octahedral sites, which exhibit $D_{3d}$ (or $\mathit{\bar{3}m}$) symmetry and build chains aligned along the six twofold axis of the cubic structure.\cite{octa} The number of equivalent octahedral sites in the unit cell is four, denoted hereafter as sites A, B, C and D, depending on their direction of  distortion, either [$\bar{1}$11], [11$\bar{1}$], [111] or [1$\bar{1}$1], respectively (Fig.~\ref{fig1} and Tab.~\ref{tab:coord}). During the Al to Cr substitution, the local $D_{3d}$ symmetry is retained and the relaxed Cr-site remains centrosymmetric, which indicates the absence of Cr 3\textit{d}-4\textit{p} mixing.\cite{Juhin} Hence, the K pre-edge features arise from pure electric quadrupole transitions (1\textit{s}$\rightarrow$3\textit{d}) but an experimental evidence of this is still lacking. As the Cr-site remains distorted in the same direction as for the Al-site, four equivalent relaxed sites are available for Cr. Hence, the electric dipole and electric quadrupole absorption cross-sections, for a given single crystal configuration, are expected to be different for a Cr impurity located at site A, B, C or D, since their orientations with respect to the incident beam are different.

\begin{table*}[!]
\caption{Coordinates of Cr atom and direction of site distortion for the four equivalent substitutional sites belonging to the rhombohedral unit cell. We also give the coordinates of the twelve other sites, obtained from the previous by the three translations of the \textit{fcc} lattice (see text and Fig.\ref{fig1}).}
\label{tab:coord}
\begin{ruledtabular}
\begin{tabular}{cccc}
site identification &  direction of site distortion & Cr-coordinates in rhombohedral unit cell  & Cr-coordinates in cubic cell  \\  \hline \\
A  &  [$\bar{1}$11] & ($\frac{1}{2}$,$\frac{1}{4}$,$\frac{1}{4}$) & ($\frac{1}{2}$,$\frac{3}{4}$,$\frac{3}{4}$) , (0,$\frac{3}{4}$,$\frac{1}{4}$) , (0,$\frac{1}{4}$,$\frac{3}{4}$)   \\ 
B  &  [11$\bar{1}$] & ($\frac{1}{4}$,$\frac{1}{4}$,$\frac{1}{2}$) & ($\frac{3}{4}$,$\frac{3}{4}$,$\frac{1}{2}$) , ($\frac{1}{4}$,$\frac{3}{4}$,0) , ($\frac{3}{4}$,$\frac{1}{4}$,0) \\ 
C  &  [111] & ($\frac{1}{2}$,$\frac{1}{2}$,$\frac{1}{2}$)& (0,0,$\frac{1}{2}$),  (0,$\frac{1}{2}$,0) , ($\frac{1}{2}$,0,0)   
\\ D  &  [1$\bar{1}$1] & ($\frac{1}{4}$,$\frac{1}{2}$,$\frac{1}{4}$) & ($\frac{3}{4}$,$\frac{1}{2}$,$\frac{3}{4}$) , ($\frac{1}{4}$,0,$\frac{3}{4}$) , ($\frac{3}{4}$,0,$\frac{1}{4}$)  \\
\end{tabular}
\end{ruledtabular}
\end{table*}

In this paper, we compare the experimental angular dependence of the Cr K pre-edge MgAl$_{2}$O$_{4}$:Cr$^{3+}$ to those obtained by theoretical calculations, combining a monoelectronic approach based on density functional theory (DFT) and multielectronic methods based on the ligand field multiplet theory (LFM). The monoelectronic approach is usually dedicated to the study of delocalized final states (e.g., the calculation of K-edge spectra) but has also provided satisfactory results for the study of Ti K pre-edge in TiO$_{2}$ and SrTiO$_{3}$,\cite{Joly,Cabaret,Yamamoto,Woicik} and also for the study of Fe K pre-edge in FeS$_{2}$.\cite{Cabaret2}  The multielectronic approach, usually dedicated to the study of localized final states (e.g., K pre-edge, L$_{2,3}$ edges of 3\textit{d} transition elements) has been succesfully applied to the case of K pre-edge in several systems.\cite{Arrio,Westre} 
Our aim is to determine the factors (site distortion, electronic interactions) prevailing at the angular dependence of Cr K pre-edge in spinel and to provide a comparison between the monoelectronic and multielectronic approaches. We also present a powerful method, based on symmetry considerations, to reduce the number of calculations needed to reconstruct the angular dependence of the cubic crystal from that of a single site. 
The paper is organized as follows. Section II is dedicated to the experimental work, including the sample description, the X-ray absorption measurements and analysis. Section III is devoted to the computational work, including the theoretical framework (Sec. III A), the details of DFT calculations (Sec. III B) and of the multiplet calculations (Sec. III C). Results are presented in Sec. IV and discussed in Sec. V.

\section{Experiments}
A natural gem-quality red spinel single crystal from Mogok (Burma), with composition (Mg$_{0.95}$Fe$_{0.01}$)$_{0.96}$(Al$_{2.02}$Cr$_{0.01}$)$_{2.03}$O$_{4}$, was investigated (for details, see Ref. \onlinecite{Juhin}). The single crystal was cut along the (110) plane (plotted in red on Fig.\ref{fig1}) and orientated according to the Laue method.\\ 
Cr K-edge (5989 eV) XAS spectra were collected at room temperature at beamline ID26 of the European Synchrotron Radiation Facility (Grenoble, France).\cite{Gauthier} The energy of the incident radiation was selected using a pair of He-cooled Si crystals with (111) orientation. The spot size on the sample was approximatively 250$\times$50~$\mathrm{\mu}$m$^{2}$. The orientated sample was placed on a rotating holder at 45$^{\circ}$ with respect to the incident beam, and turned around the [110] direction from a rotation angle $\alpha_{rot}$. The starting configuration ($\alpha_{rot}$~=~0$^{\circ}$) corresponds to $\bm{\mathrm{\hat{\varepsilon}}}$~=~[010] and $\bm{\mathrm{\hat{k}}}$~=~[$\bar{1}$00] (see Fig.\ref{fig1}). The ($\alpha_{rot}$~=~90$^{\circ}$) configuration corresponds to $\bm{\mathrm{\hat{\varepsilon}}}$~=~[$\frac{1}{2}$,$\frac{1}{2}$,$\frac{1}{\sqrt{2}}$] and $\bm{\mathrm{\hat{k}}}$~=~[-$\frac{1}{2}$,-$\frac{1}{2}$,$\frac{1}{\sqrt{2}}$]. For this sample cut and this experimental setup, the maximum variation effect is obtained by substracting the absorption recorded for $\alpha_{rot}$ = 0$^{\circ}$ from that recorded for $\alpha_{rot}$~=~90$^{\circ}$. 
One spectrum was recorded every 15$^{\circ}$ from $\alpha_{rot}$~=~0$^{\circ}$ to $\alpha_{rot}$~=~360$^{\circ}$, which enables to reconstruct the complete angular dependence of the crystal. The absorption was measured by a photodiode fluorescence detector. For each $\alpha_{rot}$ angle, ten pre-edge spectra ranging from 5987 to 5998~eV were recorded with an energy step of 0.05~eV and averaged. Two additional scans were recorded between 5985 and 6035~eV  by step of 0.2~eV, in order to merge the pre-edge on the XANES spectrum, and two more spectra were recorded between 5950 and 6350~eV by step of 0.5~eV, in order to normalize the XANES to the K-edge jump far from the edge. Self-absorption effects are negligible, because of the low Cr-content of the sample.
\section{Theory}
In this section, we recall the general expressions of the electric dipole and quadrupole absorption cross-sections  for a cubic crystal and for a site with $D_{3d}$ symmetry (Subsec. A). Then, we use the general method described in Ref.~\onlinecite{Brouder2} to calculate the angular dependence of the cubic crystal from that of a single site. This framework is illustrated in the particular case of spinel. Finally, we report the details of the monoelectronic and multielectronic calculations performed for substitutional Cr in spinel (Subsec. B and C).

\subsection{Theoretical framework}

\subsubsection{Absorption cross-sections for a cubic crystal}

The total absorption cross-section for a crystal (cubic or non-cubic), $\sigma$, is expressed as: 
\begin{equation}
\label{cross_section_cube}
\sigma(\bm{\mathrm{\hat{\varepsilon}}},\bm{\mathrm{\hat{k}}})= \sigma^{D}(\bm{\mathrm{\hat{\varepsilon}}}) + \sigma^{Q}(\bm{\mathrm{\hat{\varepsilon}}},\bm{\mathrm{\hat{k}}})
\end{equation}
where $\sigma^{D}$ is the electric dipole cross-section and $\sigma^{Q}$ is the electric quadrupole cross-section. The expression given above is valid in the absence of coupling between the electric dipole and the electric quadrupole terms: this condition is fulfilled if the system is either centrosymmetric or if, at the same time, the system is non-magnetic (no net magnetic moment on the absorbing ion) and one uses exclusively linear polarization. For Cr in MgAl$_{2}$O$_{4}$, the two types of conditions are satisfied. The dipole and quadrupole cross-sections can be expressed in function of spherical tensor components, respectively ($\sigma^{D}$(0,0), $\sigma^{D}$(2,\textit{m})) and ($\sigma^{Q}$(0,0), $\sigma^{Q}$(2,\textit{m}), $\sigma^{Q}$(4,\textit{m})), which transform under rotation like the corresponding spherical harmonics (Y$_{0}^{0}$, Y$_{2}^{m}$ and Y$_{4}^{m}$).\cite{Brouder} The tensor components are functions of $\hbar\omega$, omitted for clarity in this paper. The symmetries of the crystal restrict the possible values of $\sigma^{D}$(2,\textit{m}) and $\sigma^{Q}$(4,\textit{m}), as will be precised hereafter for the cubic case. 

The electric dipole cross-section for a cubic crystal, $\sigma_{cub}^{D}$, is isotropic (e.g., it does not depend on the direction of the polarization vector) and is equal to $\sigma_{cub}^{D}$(0,0):\cite{Brouder}
\begin{equation}
\sigma_{cub}^{D}(\bm{\mathrm{\hat{\varepsilon}}}) = \sigma_{cub}^{D}(0,0).
\end{equation}
 
The electric quadrupole cross-section for a cubic crystal, $\sigma_{cub}^{Q}$, is expressed, according to group theory (Appendix A), as: 
\begin{equation}
\label{cross_section_quad}
\sigma_{cub}^{Q}(\bm{\mathrm{\hat{\varepsilon}}},\bm{\mathrm{\hat{k}}})= \sigma_{cub}^{Q}(0,0) +\frac{20}{\sqrt{14}}(\varepsilon_{x}^{2} k_{x}^{2} +\varepsilon_{y}^{2} k_{y}^{2} + \varepsilon_{z}^{2} k_{z}^{2}-\frac{1}{5} )\sigma_{cub}^{Q}(4,0),
\end{equation} where $\sigma_{cub}^{Q}(0,0)$ is the isotropic electric quadrupole cross-section, and $\sigma_{cub}^{Q}(4,0)$ is a purely anisotropic electric quadrupole term. The polarization unit vector $\bm{\mathrm{\hat{\varepsilon}}}$ and the wave unit vector $\bm{\mathrm{\hat{k}}}$ have their coordinates expressed in the Cartesian reference frame of the cube. 

\subsubsection{Absorption cross-sections for a site with $D_{3d}$ symmetry\\}
For a site with $D_{3d}$ symmetry, the reference frame is  chosen consistently with the symmetry operations of the point group, i.e. with the \textit{z}-axis parallel to the \textit{C$_{3}$} axis of the $D_{3d}$  group.\cite{Butler} The polarization and the wave unit vectors are expressed as: 
$\bm{\mathrm{\hat{\varepsilon}}}$ =
$\begin{pmatrix}
\sin\theta~\cos\phi \\
\sin\theta~\sin\phi \\
\cos\theta
\end{pmatrix}$\\
and $\bm{\mathrm{\hat{k}}}$ =
$\begin{pmatrix}
\cos\theta~\cos\phi~\cos\psi - \sin\phi~\sin\psi \\
\cos\theta~\sin\phi~\cos\psi + \cos\phi~\sin\psi\\
-\sin\theta~\cos\psi 
\end{pmatrix}$\\\\
Hence, $\theta$, which appears in the expression of $\bm{\mathrm{\hat{\varepsilon}}}$ and $\bm{\mathrm{\hat{k}}}$, is the angle between $\bm{\mathrm{\hat{\varepsilon}}}$ and the \textit{C$_{3}$} axis.

The electric dipole absorption cross-section in $D_{3d}$ is given by:\cite{Brouder}  
\begin{equation}
\label{dipD3d}
\sigma_{D_{3d}}^{D} (\bm{\mathrm{\hat{\varepsilon}}}) = \sigma_{D_{3d}}^{D}(0,0)-\frac{1}{\sqrt{2}}(3 \cos^{2}\theta -1)\sigma_{D_{3d}}^{D}(2,0).
\end{equation} 
In order to determine $\sigma_{D_{3d}}^{D}(\bm{\mathrm{\hat{\varepsilon}}}$) for any experimental configuration ($\bm{\mathrm{\hat{\varepsilon}}}$), one needs first to determine $\sigma_{D_{3d}}^{D}(0,0)$ and $\sigma_{D_{3d}}^{D}(2,0)$, for example by performing calculations for at least two independent orientations of $\bm{\mathrm{\hat{\varepsilon}}}$. The isotropic term, $\sigma_{D_{3d}}^{D}(0,0)$, can be calculated directly by choosing $\theta$ = $\arccos \frac{1}{\sqrt{3}}$ .  \\

The electric quadrupole absorption cross-section in $D_{3d}$ is given by:\cite{Brouder}  
\begin{align}
\label{cross_section_D3d}
\nonumber&\sigma_{D_{3d}}^{Q}(\bm{\mathrm{\hat{\varepsilon}}},\bm{\mathrm{\hat{k}}})\\
\nonumber& =  \sigma_{D_{3d}}^{Q}(0,0)+\sqrt{\frac{5}{14}}(3~\sin^{2}\theta~\sin^{2}\psi -1)~\sigma_{D_{3d}}^{Q}(2,0) \\
\nonumber &+ \frac{1}{\sqrt{14}}(35~\sin^{2}\theta~\cos^{2}\theta~\cos^{2}\psi \\
\nonumber & +5\sin^{2}\theta~\sin^{2}\psi-4)~\sigma_{D_{3d}}^{Q}(4,0) \\
\nonumber&-\sqrt{10}~\sin\theta[(2\cos^{2}\theta~\cos^{2}\psi-1)\cos\theta \cos3\phi\\
& -(3~\cos^{2}\theta-1)~\sin\psi~\cos\psi~\sin3\phi]~\sigma_{D_{3d}}^{Q}(4,3).
\end{align}

To determine $\sigma_{D_{3d}}^{Q}(\bm{\mathrm{\hat{\varepsilon}}},\bm{\mathrm{\hat{k}}}$) for any experimental configuration ($\bm{\mathrm{\hat{\varepsilon}}}$,$\bm{\mathrm{\hat{k}}}$), one needs first to determine $\sigma_{D_{3d}}^{Q}(0,0)$, $\sigma_{D_{3d}}^{Q}(2,0)$, $\sigma_{D_{3d}}^{Q}(4,0)$ and $\sigma_{D_{3d}}^{Q}(4,3)$, for example by performing calculations for at least four independent orientations ($\bm{\mathrm{\hat{\varepsilon}}}$,$\bm{\mathrm{\hat{k}}}$).

\subsubsection{From a single site $D_{3d}$ to the cubic crystal\\}
In order to reconstruct the angular dependence of the cubic crystal from that of a single site with $D_{3d}$ symmetry, the tensor components have to be averaged over the equivalent sites of the cubic cell. For the electric dipole cross-section, we need a relation between ($\sigma_{D_{3d}}^{D}$(2,0), $\sigma_{D_{3d}}^{D}(0,0)$) and $\sigma_{cub}^{D}$(0,0), and for the electric quadrupole cross-section, we need a relation between ($\sigma_{D_{3d}}^{Q}$(0,0), $\sigma_{D_{3d}}^{Q}$(2,0), $\sigma_{D_{3d}}^{Q}$(4,0), $\sigma_{D_{3d}}^{Q}$(4,3)) and ($\sigma_{cub}^{Q}$(0,0), $\sigma_{cub}^{Q}$(4,0)). To do so, we have used the formulas given in Ref. \onlinecite{Brouder2}, which have been obtained from a spherical tensor analysis. This general method uses the symmetry operations of the crystal, which exchange the equivalent sites of the cubic cell, and is here illustrated in the case of spinel.  

The averages over the four equivalent sites are given by:\cite{Brouder2} 
\begin{align}
&\sigma_{cub}^{D}(0,0) =  \sigma_{D_{3d}}^{D}(0,0),\\
&\sigma_{cub}^{D}(2,0) =  0
\end{align} 

Similarly, we have: \cite{Brouder2}
\begin{eqnarray}
\label{moyenne1}
\sigma_{cub}^{Q}(0,0) &= & \sigma_{D_{3d}}^{Q}(0,0),\\
\label{moyenne2}
\sigma_{cub}^{Q}(4,0) &=&-\frac{1}{18}(7\sigma_{D_{3d}}^{Q}(4,0)+2{\sqrt{70}}~\sigma_{D_{3d}}^{Q}(4,3))
\end{eqnarray} 

\subsubsection{Calculation of the absorption cross-sections for the experimental orientations}
The electric dipole isotropic cross-section of the cubic crystal, $\sigma_{cub}^{D}$($\bm{\mathrm{\hat{\varepsilon}}}$), does not depend on the direction of the incident polarization vector $\bm{\mathrm{\hat{\varepsilon}}}$. Hence, it will be the same for every experimental configuration: 
\begin{eqnarray}
\label{alpha2}
\sigma_{cub}^{D}(\alpha_{rot})  & =  \sigma_{cub}^{D}(0,0). 
\end{eqnarray}\\
For the sample cut and the experimental setup used in this study, the expression of the electric quadrupole cross-section of the cubic crystal, is given in function of the rotation angle $\alpha_{rot}$ by:
\begin{eqnarray}
\label{alpha}
\nonumber \sigma_{cub}^{Q}(\alpha_{rot})  & =  \sigma_{cub}^{Q}(0,0) + \frac{1}{16\sqrt{14}}[-19-60\cos(2\alpha_{rot})\\
  & + 15\cos(4\alpha_{rot})] \sigma_{cub}^{Q}(4,0).
\end{eqnarray}\\
The connection between Eqs.~\ref{alpha} and \ref{cross_section_quad} is made following the definition of $\bm{\mathrm{\hat{\varepsilon}}}$ and $\bm{\mathrm{\hat{k}}}$ as functions of $\alpha_{rot}$ (Appendix B).
Eq. \ref{alpha} shows that the total angular dependence of the cubic crystal is a $\pi$-periodic function. 
The fact that the rotation axis might not be perfectly aligned with the x-ray beam or that the sample might not be perfectly homogeneous, could have introduced an additional 2$\pi$ periodic component. This component would be removed from the signal, using a filtering algorithm, based on the angular dependence recorded from 0$^{\circ}$ to 360$^{\circ}$.\cite{Cabaret2} In our experiments, this 2$\pi$ periodic-component was measured to be very small, and no filtering was applied.  \\

For the present sample cut and experimental setup, the maximum variation of the electric quadrupole cross-section is expected between $\alpha_{rot}$ = 0$^{\circ}$ and $\alpha_{rot}$ = 90$^{\circ}$. 
\begin{itemize}
\item For $\alpha_{rot}$ = 0$^{\circ}$:
\begin{align}                                              
\label{0}
\sigma_{cub}^{Q}(\alpha_{rot} = 0^{\circ}) = \sigma_{0}^{Q} -\frac{4}{\sqrt{14}} \sigma_{cub}^{Q}(4,0).
\end{align}
\item For $\alpha_{rot}$ = 90$^{\circ}$: 
\begin{align}
\label{90}
\sigma_{cub}^{Q}(\alpha_{rot} = 90^{\circ}) = \sigma_{0}^{Q} +\frac{7}{2\sqrt{14}} \sigma_{cub}^{Q}(4,0).
\end{align}
\item The isotropic cross-section is : 
\begin{align}
\label{iso}
\nonumber \sigma_{iso}^{Q} & = \frac{1}{15}(8~\sigma_{cub}^{Q}(\alpha_{rot} = 90^{\circ})+7~\sigma_{cub}^{Q}(\alpha_{rot}  = 0^{\circ}))\\
& =\sigma_{cub}^{Q}(0,0).
\end{align}
\item The dichroic term is :
\begin{align}
\label{dichro}
\nonumber \sigma_{dichro}^{Q} & =  \sigma_{cub}^{Q}(\alpha_{ rot} = 90^{\circ})-\sigma_{cub}^{Q}(\alpha_{rot} = 0^{\circ})\\
& = \frac{15}{2\sqrt{14}}\sigma_{cub}^{Q}(4,0).
\end{align}
\end{itemize}

\subsection{Computational details}

\subsubsection{Density Functional Theory Calculations}\

The computation of the electric dipole and electric quadrupole absorption cross-sections were done using a first-principles total energy code based on DFT in the Local Density Approximation with spin-polarization (LSDA).\cite{Paratec} We used periodic boundary conditions, plane wave basis set and norm conserving pseudopotentials\cite{Trou} in the Kleiman Bylander form.\cite{Klein} The parameters for the pseudopotential generation are given in Ref. \onlinecite{Juhin}. 

We started from a host structure of MgAl$_{2}$O$_{4}$, which is obtained by an \textit{ab initio} energy minimization calculation. In this calculation, the lattice parameter was fixed to its experimental value,\cite{Yamanaka} while the atomic positions were allowed to vary to minimize the total energy and the interatomic forces. We then relaxed a 2$\times$2$\times$2 rhomboedral supercell containing one Cr atom in substitution for Al (i.e., 1 Cr, 31 Al, 16 Mg and 64 O), with the basis vectors expressed in a cubic frame. The supercell was large enough to avoid interactions between neighboring Cr atoms. As the Cr impurity is in its high-spin state, the spin multiplet S$_{z}$=$\frac{3}{2}$ is imposed for the supercell. The atomic positions in the supercell were allowed to vary, in order to minimize the total energy and the interatomic forces. We used a 90 Ry energy cutoff and a single \textit{k}-point sampling in the Brillouin zone. The Cr-site, after relaxation, still exhibits a $D_{3d}$ symmetry, with an inversion center, one \textit{C$_{3}$} axis and three \textit{C$_{2}$} axis. 

The Cr K-edge absorption cross-section was computed using the method described in Refs.~\onlinecite{Taille,Caba}. First, we calculated self-consistently the charge density of the system, with a 1\textit{s} core-hole on the substitutional Cr atom. Then, the all-electron wave functions were reconstructed within the projector augmented wave framework.\cite{Blochl} The absorption cross-section was computed as a continued fraction, using a Lanczos basis constructed recursively.\cite{Hay,Hay2} We used a 70 Ry energy cutoff for the plane-wave expansion, one \textit{k} point for the self-consistent spin-polarized charge density calculation, and a Monkhorst-Pack grid of 3$\times$3$\times$3 \textit{k}-points in the Brillouin zone for the absorption cross-section calculation. For the convolution of the continued fraction, we used an energy-dependent broadening parameter $\gamma$, which takes into account the main photoelectron damping modes (core-hole lifetime and imaginary part of the photoelectron self-energy). The energy-dependent $\gamma$ used in this study is that of Ref.~\onlinecite{Gaudry}. The calculated spectrum was then shifted in energy to the experimental one: the maximum of absorption is set at 6008.5~eV. The absorption edge jump is set to 1, so that experimental and calculated spectra for all figures are normalized absorption. In such a way, the calculated pre-edge could be compared directly to the experimental one. 
As mentioned previously, the four substitutional sites will exhibit different spectra for the electric dipole and quadrupole operators, since their orientations are different with respect to the incident beam absorption of x-rays with given ($\bm{\mathrm{\hat{\varepsilon}}}$,$\bm{\mathrm{\hat{k}}}$). The general method to obtain the angular dependence measured for the cubic crystal is to compute the electric dipole and electric quadrupole absorption cross-sections for a Cr impurity lying in each of the four trigonally distorted sites A, B, C and D, and then to take the average. However, this heavy \textit{brute force} method requires the calculation of four monoelectronic potentials with core-hole (after previous associated structural relaxation). The number of calculations can be drastically reduced if we take advantage of the symmetry properties of the crystal, which enables to perform the calculations for only one substitutional site (site A, with coordinates of (0, $\frac{1}{4}$, $\frac{3}{4}$) and direction of distortion [$\bar{1}$11]). This method is detailed in Appendix C.\\

\begin{figure}[!]
\includegraphics[width=7.9cm]{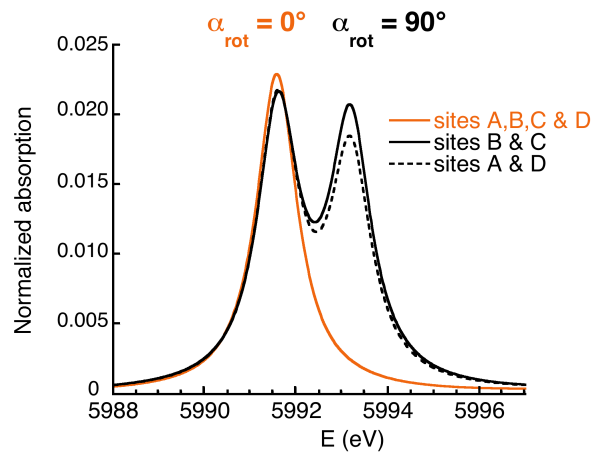}
\caption{\label{fig2} (Color online) Pre-edge spectra calculated for site A, B, C and D in the electric quadrupole approximation. The orange lines present the spectra calculated for $\alpha_{rot}$~=~0$^{\circ}$ ($\bm{\mathrm{\hat{\varepsilon}}}$~=~[010], $\bm{\mathrm{\hat{k}}}$~=~[$\bar{1}$00]), while the black lines are the spectra calculated for $\alpha_{rot}$~=~90$^{\circ}$ ($\bm{\mathrm{\hat{\varepsilon}}}$~=~[$\frac{1}{2}$,$\frac{1}{2}$,$\frac{1}{\sqrt{2}}$], $\bm{\mathrm{\hat{k}}}$~=~[-$\frac{1}{2}$,-$\frac{1}{2}$,$\frac{1}{\sqrt{2}}$]).}
\end{figure}
Figure \ref{fig2} presents the normalized electric quadrupole cross-section calculated for the four equivalent sites, for $\alpha_{rot}$~=~0$^{\circ}$ and  $\alpha_{rot}$~=~90$^{\circ}$. The spectra calculated for $\alpha_{rot}$~=~0$^{\circ}$ (orange line) are equal for the four sites. For $\alpha_{rot}$~= ~90$^{\circ}$, sites B and C give the same spectra (black solid line), as well as sites A and D (black dashed line). We observe a slight difference in intensity for the peak at 5993.2~eV: this is indeed a consequence of the fact that sites (A, D) and (B, C)  have different orientations with respect to the incident beam, and that their symmetry differs from \textit{O}$_{h}$. Because the trigonal distortion of the octahedra is small in spinel, the anisotropic behaviour of the sites is limited for the investigated configurations. However, the effect of the trigonal environment can have drastic consequences when the distortion is more pronounced.

\subsubsection{Ligand Field Multiplet Calculations}
In order to extract quantitative information from the angular dependence of the pre-edge, we have performed LFM calculations using the method developed by T. Thole in the framework established by Cowan and Butler.\cite {Thole,Butler,Cowan} In this approach, Cr$^{3+}$ is considered as an isolated ion embedded in a crystal field potential. The band structure of the solid is not taken into account, which prevents to calculate transitions to delocalized (i.e., non-atomic) levels. In other words, the LFM approach can be used to calculate K pre-edge spectra, but the edge region cannot be computed. Since the Cr-site is centrosymmetric, no hybridization is allowed between the 3\textit{d}-orbitals and the 4\textit{p}-orbitals of Cr. Hence, the pre-edge is described by the transitions from the initial state 1\textit{s}$^{2}$3\textit{d}$^{3}$ to the final state 1\textit{s}$^{1}$3\textit{d}$^{4}$. 

We expose briefly the principles of multiplet calculations but details can be found in other references (see for example Ref.~\onlinecite{Groot}). This approach takes into account all the 3\textit{d}-3\textit{d} and 1\textit{s}-3\textit{d} electronic Coulomb interactions, as well as the spin-orbit coupling on every open shell of the absorbing atom, and treats its geometrical environment through a crystal field potential. In the electric quadrupole approximation, the spectrum is calculated as the sum of all possible transitions for an electron jumping from the 1\textit{s} level toward one 3\textit{d} level according to:
\begin{equation}
\sigma^{Q}(\bm{\mathrm{\hat{\varepsilon}}},\bm{\mathrm{\hat{k}}})= \pi^{2}~k^2~\alpha\hbar\omega\sum_{I,F}\frac{1}{d_{I}}|\langle{F}|\bm{\mathrm{\hat{\varepsilon}}}\cdot\textbf{r}\bm{\mathrm{\hat{k}}}\cdot\textbf{r}|{I}\rangle|^{2}\delta(E_{F}-E_{I}-\hbar\omega),
\end{equation} 
where $|$I$\rangle$ and $|$F$\rangle$ are the multielectronic initial and final states, of respective energies  E$_{I}$, E$_{F}$, and \textit{d}$_{I}$ the degeneracy of the initial state.\cite{footnote2}.\\

Once the $|$I$\rangle$ and $|$F$\rangle$ states have been calculated, the absolute intensities of the pre-edge spectra are calculated in \AA$^{2}$ at T~=~300 K. The population of the ground-state levels $|$I$\rangle$ is given by a Boltzmann law. The spectra are convoluted by a Lorentzian (with HWHM = 0.54~eV) and a Gaussian (with FWHM = 0.85~eV), which respectively take into account the lifetime of the 1\textit{s} core-hole for Cr and the instrumental resolution. Finally, the transitions are normalized by the edge jump at the Cr K edge, calculated for a Cr atom from Ref.~\onlinecite{Gullikson} as 4.48 10$^{-4}$ \AA$^{2}$. Hence, the calculated spectra can be directly compared to the normalized experimental ones.

The electric quadrupole absorption cross-section was calculated for a Cr$^{3+}$ ion lying in $D_{3d}$ symmetry, according to the method described above. The crystal-field parameters used in the calculation are those derived from optical absorption spectroscopy ($D_q$~=~0.226~eV, \textit{D}$_{\sigma}$~=~-0.036~eV, $D_{\tau}$~=~0.089~eV).\cite{Wood} We used the scaling factor of the Slater integrals ($\kappa$=~0.7), related to B and C Racah parameters, given in the same reference. The only adjustable parameter is the absolute position in energy. 

As mentioned in Sec. III A 2 (Eq.~ \ref{cross_section_D3d}), one needs first to determine $\sigma_{D_{3d}}^{Q}(0,0)$, $\sigma_{D_{3d}}^{Q}(2,0)$, $\sigma_{D_{3d}}^{Q}(4,0)$ and $\sigma_{D_{3d}}^{Q}(4,3)$, in order to determine $\sigma_{cub}^{Q}(0,0)$ and $\sigma_{cub}^{Q}(4,0)$  using Eqs~\ref{moyenne1} ~and~\ref{moyenne2}. This is done by performing four multiplet calculations, which provide four independent values of the electric quadrupole cross-section (Appendix D). Once this first step has been performed, we used Eq. \ref{moyenne1} and \ref{moyenne2} to derive $\sigma_{cub}^{Q}(0,0)$ and $\sigma_{cub}^{Q}(4,0)$. The electric quadrupole cross-section of the cubic crystal can then be calculated for any experimental configuration using Eq.~\ref{alpha}. 
The electric quadrupole cross-section for $\alpha_{rot}$~=~0$^{\circ}$ and $\alpha_{rot}$~=~ 90$^{\circ}$, the dichroic and the isotropic spectra were determined respectively according to Eqs~\ref{0}-\ref{dichro}.

\section{Results}
\subsection{DFT calculations}

\subsubsection{Comparison with experiment}\

\begin{figure}[!b]
\includegraphics[width=7.9cm]{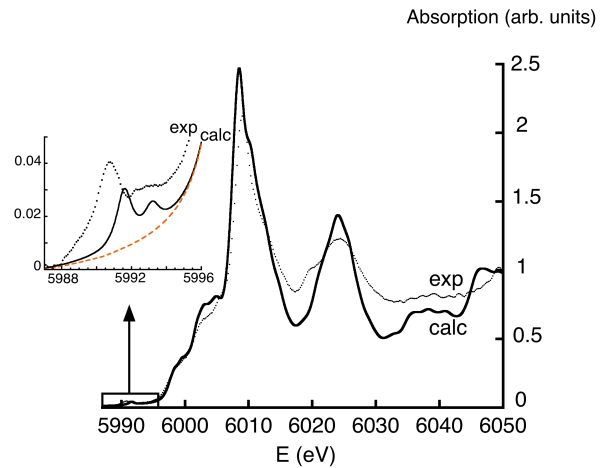}
\caption{\label{fig3} (Color online) Comparison between experimental (dotted line) and calculated (solid line) isotropic XANES spectra at the Cr K-edge in spinel. The calculation was performed using DFT-LSDA (see Sec. III B1). The inset presents the spectra in the pre-edge region. The dashed orange line is the calculated electric dipole contribution.}
\end{figure}
The XANES spectrum, calculated for the cubic crystal by first-principles calculations (solid line), is shown in Fig.~\ref{fig3} and compared to the experimental spectrum (dotted line). As we mentioned above, the main absorption edge is due to electric dipole transitions. Hence, the XANES spectrum does not show any angular dependence, except in the pre-edge region. The agreement between the experimental and theoretical spectra is good, since all the features are reproduced by the calculation. A more detailed discussion is reported in Ref.~\onlinecite{Juhin}. The inset of Fig.~\ref{fig3} shows the theoretical isotropic XANES spectrum in the pre-edge region (black solid line). This spectrum is the sum of the isotropic electric dipole (orange dashed line) and the electric quadrupole contributions. Our calculations show that electric dipole transitions do not contribute to the pre-edge, except by a background, which is actually the tail of the absorption edge (1\textit{s}$\rightarrow$\textit{p} transitions). This is a clear confirmation that Cr K pre-edge features are due to a pure electric quadrupole contribution.
In the pre-edge region, the calculated isotropic spectrum is in satisfactory agreement with experiment, since the two features visible in the pre-edge are reproduced. Similar calculations have been successfully performed to calculate the K pre-edge for substitutional Cr$^{3+}$ in corundum and beryl,\cite{Gaudry,Gaudry2} with a good agreement between the experimental and theoretical data. This shows that a monoelectronic approach can reproduce pre-edge features, as can be measured on powder spectra. However, the position of the theoretical spectrum is shifted by about 0.9~eV relative to experiment. This shift, which has been already observed in several systems,\cite{Gaudry,Gaudry2,Cabaret,Cabaret2} is due the limitation of DFT-LSDA in the modelling of electron-hole interaction. In the calculation, the effect of the core-hole is to shift the 3\textit{d} levels to lower energy, with respect to the main edge. Unfortunately, this effect is not sufficient to reproduce the experimental data, because the core-hole seems to be partly screened.\cite{Joly} This could be improved by taking into account the self-energy of the photoelectron.\cite{Natoli} \\

\begin{figure}[!]
\includegraphics[width=7.9cm]{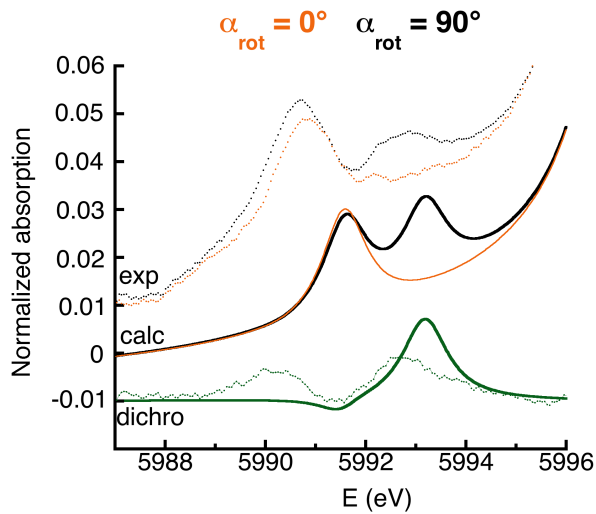}
\caption{\label{fig4} (Color online) Comparison between experimental (dotted line) and calculated (solid line) Cr K-pre edge spectra in spinel, using DFT-LSDA. The orange lines correspond to $\bm{\mathrm{\hat{\varepsilon}}}$ =[010], $\bm{\mathrm{\hat{k}}}$~=~[$\bar{1}$00] ($\alpha_{rot}$=0$^{\circ}$). The black lines correspond to $\bm{\mathrm{\hat{\varepsilon}}}$~=~[$\frac{1}{2}$,$\frac{1}{2}$,$\frac{1}{\sqrt{2}}$], $\bm{\mathrm{\hat{k}}}$~=~[-$\frac{1}{2}$,-$\frac{1}{2}$,$\frac{1}{\sqrt{2}}$] ($\alpha_{rot}$~=~90$^{\circ}$). The dichroic signal (green lines) is the difference between the black and red lines, for the experimental (dotted) and the calculated spectra (solid), respectively.}
\end{figure}
The experimental and calculated pre-edge spectra for the two configurations, which give the maximum dichroic signal for the sample cut, are shown in Fig.~\ref{fig4}. The number of peaks is well reproduced in both cases by the calculation. For both spectra, the intensity of the first peak at about 5990.7~eV is close to 4~\% of the absorption edge on the experimental data, but underestimated by 25~\% in the calculation. The relative intensity of the peak at about 5992.7~eV is overestimated in the 90$^{\circ}$ configuration. Additionally, the energy splitting between the two peaks is underestimated by the calculation (1.6~eV vs 2.0~eV experimentally). The small energy shift of the first peak between the two configurations, observed as positive in the experimental data, is calculated as negative. As a consequence of those several discrepancies, the theoretical dichroic signal is not in good agreement with the experimental one. Compared to the Ti K pre-edge calculations in rutile and SrTiO$_3$, using a similar monoelectronic approach and reported in several studies,\cite{Joly,Cabaret,Yamamoto,Woicik} the significant discrepancy observed for Cr in spinel may seem at first sight unexpected. However, we underline the fact that interlectronic repulsions become crucial for localized final states (e.g., for 1\textit{s}$\rightarrow$3\textit{d} transitions), and that Ti has no \textit{d} electrons in the systems studied. This shows that the electronic interactions on the Cr atom are too significant to reproduce quantitatively the angular dependence of Cr-spinel in a monoelectronic approach, although the average description (i.e., the isotropic spectrum) is satisfactory.

Nevertheless, the monoelectronic calculation is able to reproduce the correct number of peaks. Since this monoelectronic approach does not take into account spin-orbit coupling and does not fully describe the 3\textit{d}-3\textit{d} electronic repulsion, a monoelectronic chemical vision of an isolated Cr$^{3+}$ ion can be applied for the interpretation of the calculated features. 
 
\subsubsection{Assignment of the calculated monoelectronic transitions within an atomic picture}\
\begin{figure}[!]
\includegraphics[height=6.2cm]{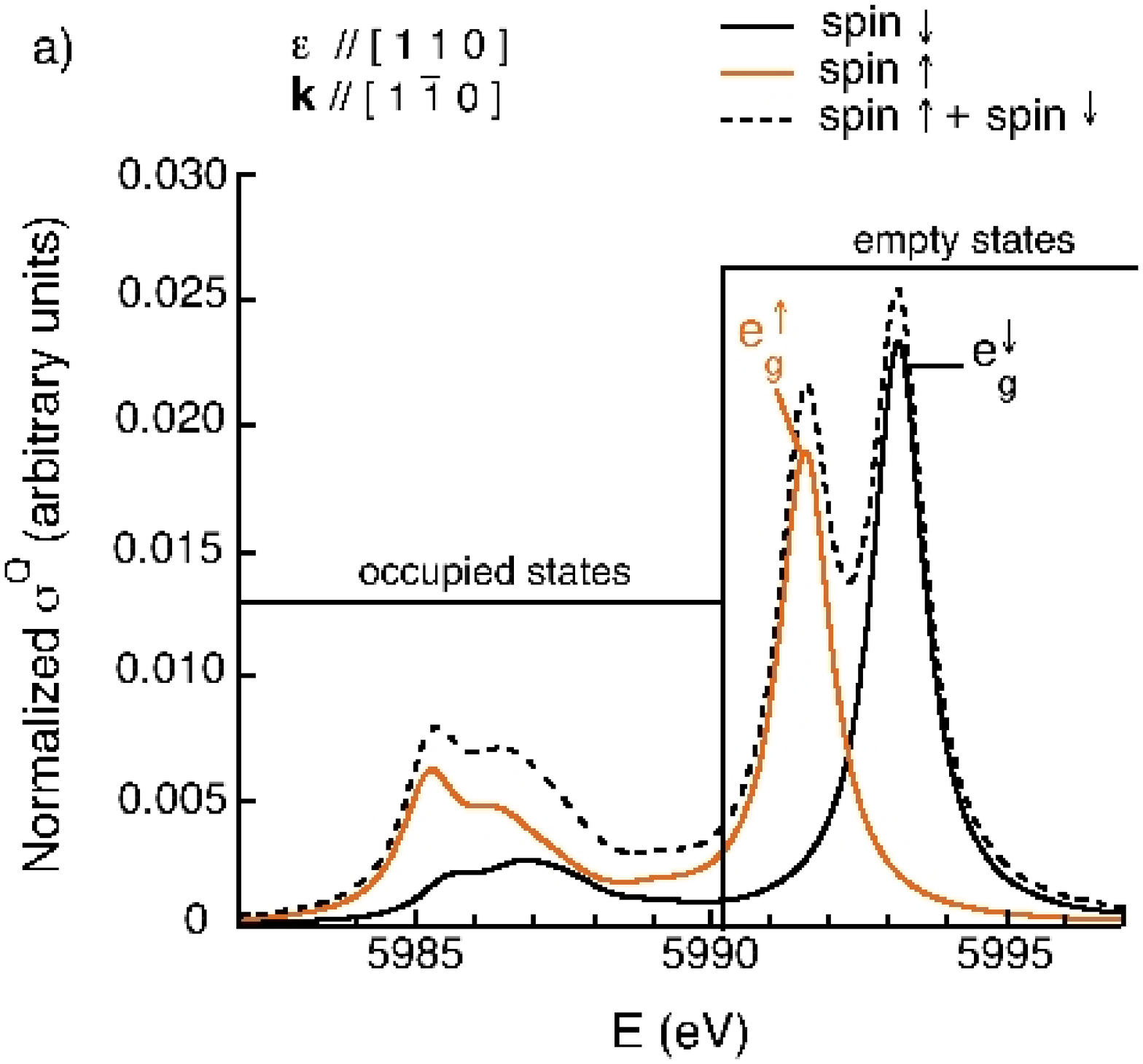}
\includegraphics[height=6.2cm]{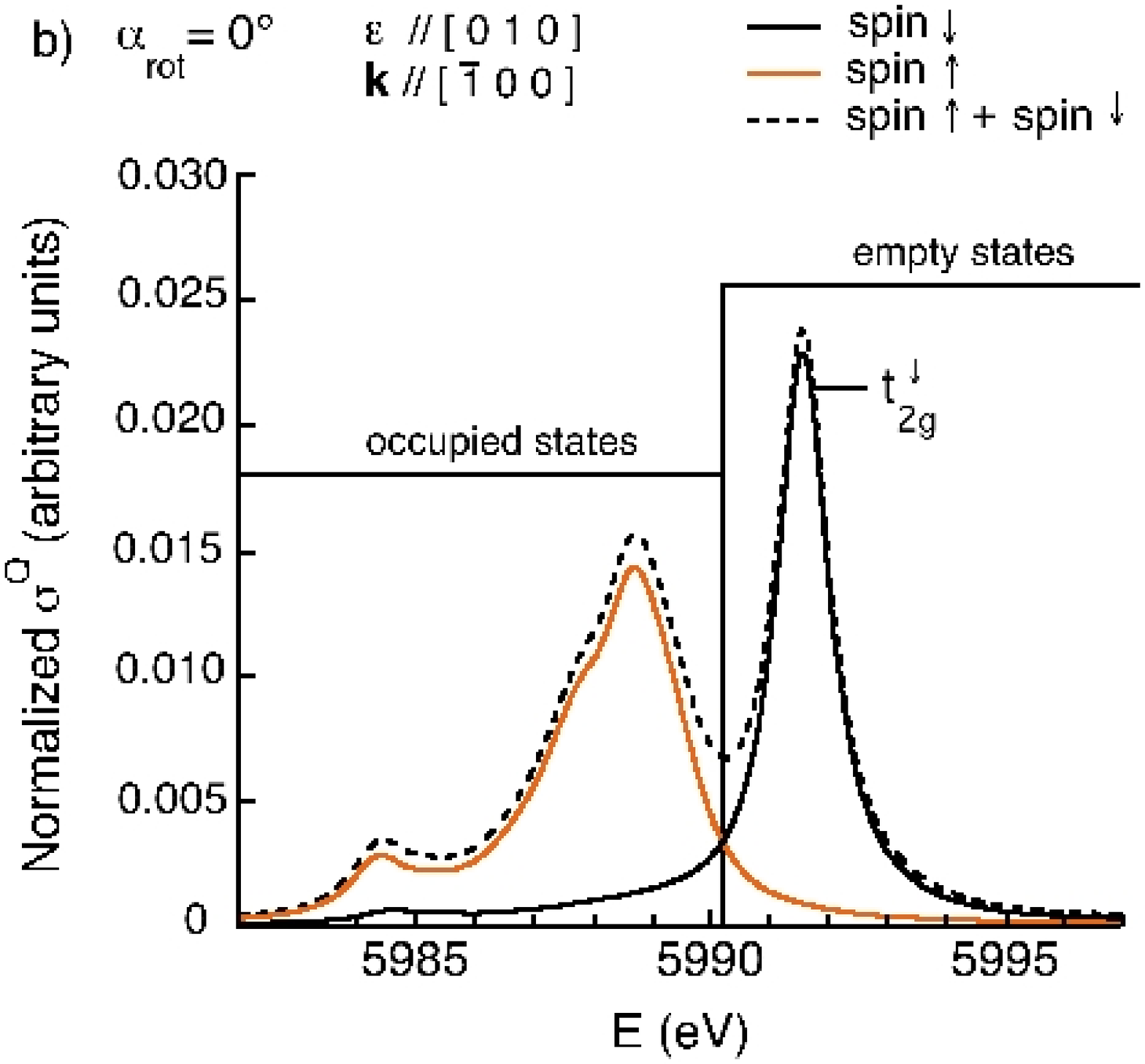}
\includegraphics[height=6.2cm]{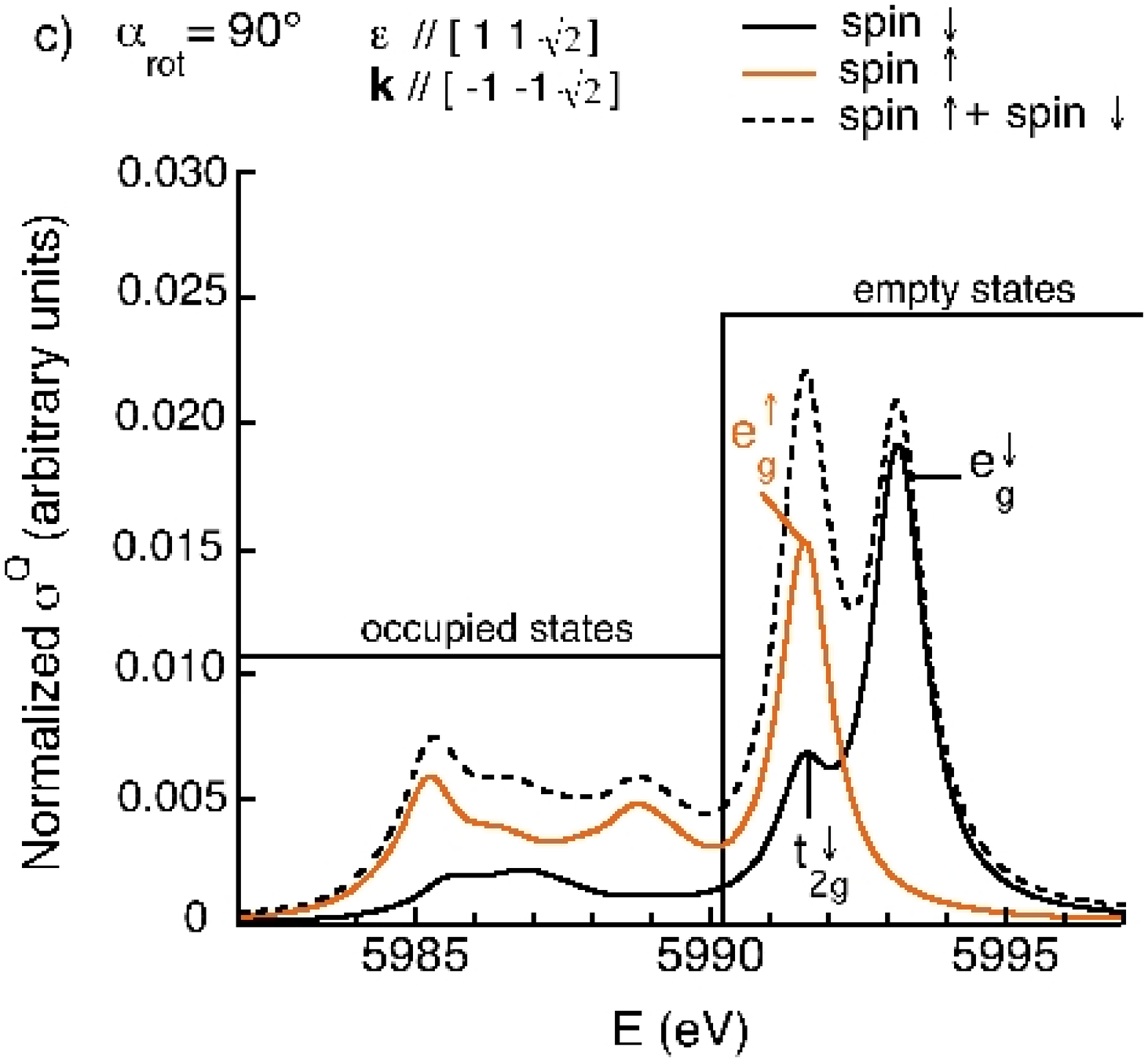}
\caption{\label{fig5} (Color online) Electric quadrupole transitions calculated for site A, using spin-polarization. The upper panel (a) presents the spectrum calculated for ($\bm{\mathrm{\hat{\varepsilon}}}$~=~[$\frac{1}{\sqrt{2}}$,$\frac{1}{\sqrt{2}}$,0], $\bm{\mathrm{\hat{k}}}$~=~[$\frac{1}{\sqrt{2}}$,-$\frac{1}{\sqrt{2}}$,0]). The middle panel (b) presents the spectrum calculated for $\alpha_{rot}$~=~0$^{\circ}$ ($\bm{\mathrm{\hat{\varepsilon}}}$~=~[010], $\bm{\mathrm{\hat{k}}}$~=~[$\bar{1}$00]). The lower panel (c) presents the spectrum calculated for $\alpha_{rot}$~=~90$^{\circ}$ ($\bm{\mathrm{\hat{\varepsilon}}}$~=~[$\frac{1}{2}$,$\frac{1}{2}$,$\frac{1}{\sqrt{2}}$], $\bm{\mathrm{\hat{k}}}$~=~[-$\frac{1}{2}$,-$\frac{1}{2}$,$\frac{1}{\sqrt{2}}$]). For each configuration, the respective contributions of the two spins are plotted (black solid line for spin down, orange solid line for spin up), as well as the sum (dashed line). The Fermi level is located approximately around 5990~eV.}
\end{figure}
In the following, we shall concentrate on the spectra associated to the local symmetry (i.e., calculated for site A). In the monoelectronic calculations, the spin multiplet S$_{z}$=$\frac{3}{2}$ is imposed for the supercell, since the Cr impurity is in its high-spin state. Cr$^{3+}$ has an initial electronic configuration ($t_{2g}^{\uparrow})^{3}$($e_{g})^{0}$, which means that Cr$^{3+}$ is a fully magnetized paramagnetic ion in the calculation, while it is paramagnetic in the experiment. Indeed, it is not possible to impose the fourfold degenerate S=$\frac{3}{2}$ ground-state in the DFT calculation, that requires non-degenerate ground states. In order to assign the transitions visible in the experimental spectra, we would need to calculate the average of the spectra for S$_z$=$\frac{3}{2}$, S$_z$=-$\frac{3}{2}$, S$_z$=$\frac{1}{2}$ and S$_z$=-$\frac{1}{2}$. However, it is not possible to do the calculation for S$_z$=$\pm\frac{1}{2}$ in the Kohn-Sham formalism, since they are linear combinations of three Slater determinants.\cite{Slater} Nevertheless, the spin-polarized computation of the XANES spectrum for S$_z$=$\frac{3}{2}$ enables to understand the origin of the pre-edge features: the contribution of the two spins ($\uparrow$ and $\downarrow$) can be indeed separated, which means that we can deduce whether the 3\textit{d}-orbitals have been reached by a 1\textit{s} electron with spin $\uparrow$ or $\downarrow$, and this for different expressions of the electric quadrupole operator.

As shown in Fig.~\ref{fig1}, although the distortion of the octahedra has been slighlty exagerated, the oxygen ligands are located approximately along the fourfold axis of the cube for all the equivalent sites A, B, C and D. Thus, the analysis made for site A provides an assignment, which is also valid (mutatis mutandis) for the equivalent sites. For a given configuration ($\bm{\mathrm{\hat{\varepsilon}}}$, $\bm{\mathrm{\hat{k}}}$), we can easily deduce from the expression of the electric quadrupole operator which 3\textit{d}-orbital has been probed in this transition. The interpretation of the features is possible through group theory in the monoelectronic approach, using the branching rules of \textit{O}$_{h}$ $\supset$ $D_{3d}$ (Appendix E). The \textit{d} orbitals belong to the \textit{t}$_{2g}$(\textit{O}$_{h}$) and \textit{e}$_{g}$(\textit{O}$_{h}$) irreducible representations within octahedral symmetry. When lowering the symmetry to $D_{3d}$, the \textit{t}$_{2g}$(\textit{O}$_{h}$) irreducible representation is split into the \textit{e}$_{g}$($D_{3d}$) and \textit{a}$_{1g}$($D_{3d}$) irreducible representations. To indicate that they come from \textit{t}$_{2g}$(\textit{O}$_{h}$), they will be written as \textit{e}$_{g}$($t_{2g}$) and \textit{a}$_{1g}$($t_{2g}$). The \textit{e}$_{g}$(\textit{O}$_{h}$) irreducible representation becomes the \textit{e}$_{g}$($D_{3d}$) irreducible representation, designed hereafter as \textit{e}$_{g}$($e_{g}$). 
 
The normalized electric quadrupole cross-sections calculated for site A are shown in Fig.~\ref{fig5} for three different configurations ($\bm{\mathrm{\hat{\varepsilon}}}$, $\bm{\mathrm{\hat{k}}}$). For a better understanding of the structures, the electric quadrupole transitions to both occupied and empty states are represented. For  ($\bm{\mathrm{\hat{\varepsilon}}}$~=~[$\frac{1}{\sqrt{2}}$,$\frac{1}{\sqrt{2}}$,0], $\bm{\mathrm{\hat{k}}}$~=~[$\frac{1}{\sqrt{2}}$,-$\frac{1}{\sqrt{2}}$,0]) (Fig.~\ref{fig5}a), the electric quadrupole operator is expressed as $\hat{O}_{a}=\frac{1}{2}(x^2\!-\!y^2)$, which enables to probe the 3\textit{d} electronic density in the \textit{x$^{2}$-y$^{2}$} direction, i.e., along Cr-O bonds: the orbitals probed are the \textit{e}$_{g}$(\textit{e}$_{g}$), which are empty for spin $\uparrow$ and $\downarrow$, since they are coming from the \textit{e}$_{g}$(\textit{O}$_{h}$) levels. Fig.~\ref{fig5}a shows that, indeed, two peaks are obtained at 5991.6~eV and 5993.2~eV, above the Fermi level. Below the Fermi level  at 5990 eV, a broad structure is observed between 5982~eV and 5990~eV, which corresponds to \textit{e}$_{g}$ states hybridized with the \textit{p}-orbitals of the oxygens. \\ 
For $\alpha_{rot}$~=~0$^{\circ}$ ($\bm{\mathrm{\hat{\varepsilon}}}$~=~[010], $\bm{\mathrm{\hat{k}}}$=[$\bar{1}$00]), one single peak is obtained in the empty states at 5991.6~eV for spin $\downarrow$ (see Fig.~\ref{fig5}b, black line). For this orientation, the electric quadrupole operator, expressed as $\hat{O}_{b}=xy$, enables to probe the \textit{d} electronic density, projected on Cr, in the \textit{xy} direction, i.e., between the Cr-O bonds. The \textit{e}$_{g}$(\textit{t}$_{2g}$) and \textit{a}$_{1g}$(\textit{t}$_{2g}$) orbitals having a component along \textit{xy}, as indicated by their expressions in Appendix D (Eq.~\ref{states}), they are probed in the transition. As these states coming from the $t^{\uparrow}_{2g}$(O$_{h}$) are fully occupied, they can be reached only by a photoelectron with spin $\downarrow$. This is indeed consistent with our results. The splitting between \textit{e}$^{\downarrow}_{g}$ and \textit{a}$^{\downarrow}_{1g}$ is not visible, which is an indication of a small trigonal distortion for the Cr-site in spinel. Below the Fermi level, a broad structure with an intense peak at 5988.7~eV is observed. The intense peak corresponds to the occupied \textit{e}$^{\uparrow}_{g}$(\textit{t}$_{2g}$) levels. To interpret the origin of the broad structure, we have to remind that the \textit{e}$_{g}$(\textit{t}$_{2g}$) states can hybridize with the \textit{e}$_{g}$(\textit{e}$_{g}$) levels, since they belong to the same irreducible representation in $D_{3d}$. As mentioned previously, the hybridization of the mixed \textit{e}$_{g}$ states with the \textit{p}-orbitals of the oxygens gives rise to the structures visible below 5988 eV. \\
For $\alpha_{rot}$ = 90$^{\circ}$ ($\bm{\mathrm{\hat{\varepsilon}}}$ = [$\frac{1}{2}$,$\frac{1}{2}$,$\frac{1}{\sqrt{2}}$], $\bm{\mathrm{\hat{k}}}$ = [-$\frac{1}{2}$,-$\frac{1}{2}$,$\frac{1}{\sqrt{2}}$]) (Fig.~\ref{fig5}c), two peaks are obtained above the Fermi level. For this orientation, the electric quadrupole operator is expressed as $\hat{O}_{c}$=$\frac{3z^{2}-r^{2}}{4}\!-\!\frac{xy}{2}$, which enables to probe the 3\textit{d} electronic density both in the \textit{xy} and $3z^{2}\!-\!r^{2}$ directions. For the $3z^{2}\!-\!r^{2}$ component, the levels probed are the $e_{g}$($e_{g}$), as for the first orientation studied (Fig.~\ref{fig5}a). For the \textit{xy} component, the levels probed are the $e_{g}$($t_{2g}$) and \textit{a}$_{1g}$(\textit{t}$_{2g}$), as for the second orientation (Fig.~\ref{fig5}b). Fig.~\ref{fig5}c shows that the spectrum is a close combination of the transitions visible on the two previous spectra (Fig.~\ref{fig5}a and b), and the assignment of the structures is made clear from the two previous cases. The position of the \textit{t}$^{\downarrow}_{2g}$(\textit{e}$_{g}$) peak is close to that of the \textit{e}$^{\uparrow}_{g}$(\textit{e}$_{g}$) at 5991.6~eV. The energy difference (1.6~eV) between \textit{t}$^{\downarrow}_{2g}$(\textit{e}$_{g}$) and \textit{e}$^{\downarrow}_{g}$(\textit{e}$_{g}$)
gives an idea of the \textit{t}$^{\downarrow}_{2g}$(\textit{O}$_{h}$)-\textit{e}$^{\downarrow}_{g}$(\textit{O}$_{h}$) splitting due to the crystal field. This can be compared to the experimental crystal-field splitting (2.26~eV), derived from optical absorption spectroscopy, but one should keep in mind that the crystal field splitting in the monoelectronic picture is associated with spin $\uparrow$ levels. \\
For the configuration $\hat{O}_{c}$, the $3z^{2}\!-\!r^{2}$ component enables to probe the \textit{e}$_{g}(\textit{O}_h)$ states, as $x^{2}\!-\!y^{2}$. Considering the normalization factors in the expression of the \textit{d} orbitals, the magnitude of the transition operator along $3z^{2}\!-\!r^{2}$ is $\sqrt{3}$ times bigger than the magnitude of the transition operator along $x^{2}\!-\!y^{2}$, which is 2 times bigger than the magnitude of the transition operator along \textit{xy}. $\hat{O}_{c}$ thus appears as a linear combination of the two operators $\hat{O}_{a}$ and $\hat{O}_{b}$, with respective weights of $\frac{\sqrt{3}}{2}$ and $\frac{1}{2}$. If no coupling occurs between $\hat{O}_{a}$ and $\hat{O}_{b}$ when calculating the square matrix elements $|\langle{f}|\bm{\mathrm{\hat{\varepsilon}}}\cdot\textbf{r}\bm{\mathrm{\hat{k}}}\cdot\textbf{r}|{i}\rangle|^{2}$, the third cross-section (c) should be the linear combination of the two cross-sections (a) and (b) obtained for $\hat{O}_{a}$ and $\hat{O}_{b}$, with respective weights of $\frac{3}{4}$ and $\frac{1}{4}$. However, the linear combination of the two cross-sections (not shown in Fig.~\ref{fig5}) and the cross-section obtained for the linear combination of the transition operators are slightly different, which indicates a small interference between the \textit{xy} and $3z^{2}\!-\!r^{2}$ (or $x^{2}\!-\!y^{2}$) components. The interference is a clear evidence of the \textit{e}$^{\downarrow}_{g}$(\textit{e}$_{g}$) and \textit{e}$^{\downarrow}_{g}$(\textit{t}$_{2g}$) hybridization due to the $D_{3d}$ local symmetry.\\

\subsection{LFM calculations}

\subsubsection{Comparison with experiment}\
\begin{figure}[!b]
\includegraphics[width=7.9cm]{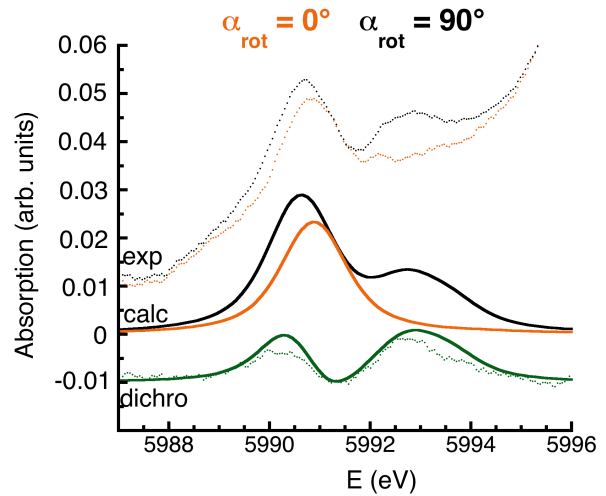}
\caption{\label{fig6}(Color online) Comparison between experimental (dotted line) and calculated (solid line) Cr K pre-edge spectra in spinel, using Ligand Field Multiplet theory in $D_{3d}$ symmetry. The orange lines correspond to $\bm{\mathrm{\hat{\varepsilon}}}$~=~[010], $\bm{\mathrm{\hat{k}}}$~=~[$\bar{1}$00] ($\alpha_{rot}$~=~0$^{\circ}$). The black lines correspond to $\bm{\mathrm{\hat{\varepsilon}}}$~=~[$\frac{1}{2}$,$\frac{1}{2}$,$\frac{1}{\sqrt{2}}$], $\bm{\mathrm{\hat{k}}}$~=~[-$\frac{1}{2}$,-$\frac{1}{2}$,$\frac{1}{\sqrt{2}}$] ($\alpha_{rot}$~=~90$^{\circ}$). The dichroic signal is the difference between the black and orange lines.}
\end{figure}
For the two experimental configurations, Fig.~\ref{fig6} presents the experimental Cr K pre-edge spectra (dotted line), the theoretical spectra obtained by LFM calculations (solid line) and the corresponding dichroic signals. The calculated pre-edges have been obtained for the cubic crystal from a calculation performed for a single site with $D_{3d}$ symmetry. For each configuration, the shape of the spectrum is well reproduced by the calculation. In the experimental data, the position of the first peak is shifted by approximately +0.15~eV for $\alpha_{rot}$~=~0$^{\circ}$, compared to that in the $\alpha_{rot}$~=~90$^{\circ}$ configuration. This relative shift is also well reproduced in the calculated spectra. For $\alpha_{rot}$~=~90$^{\circ}$, the relative intensity of the two peaks is in good agreement with the experimental data. The shape of the dichroic signal is well reproduced by the calculation: in fact, the x-ray linear dichroism of the crystal is well described in the multiplet approach, suggesting that the calculation includes the necessary multielectronic interactions on the Cr atom. We recall that the crystal-field parameters used in the calculation are those obtained from optical absorption spectroscopy (see Appendix F $\&$ G for the correspondance between the experimental crystal-field parameters and the parameters used in the multiplet calculations). However, the intensity of the dichroic signal is overestimated by 20 \% in the calculation. The first reason for this overestimation is that the calculated spectra have been normalized by the edge jump at the Cr K edge, which was calculated for an isolated Cr without considering the influence of the crystal structure according to Ref.~\onlinecite{Gullikson}. This can account for a few percent in the discrepancy. Another few percent possibly lie in the normalization of the experimental data, since we used the average of two spectra, which  were recorded between 5950~eV and 6350~eV with a rather large energy step (0.5~eV). This can introduce limited noise and thus uncertainty on the normalization. A third source of error is that the crystal-field parameters used in the calculation might be slightly different in the excited state than in the ground-state, because of the influence of the core-hole: for example, if \textit{D}$_q$ is increased by 2~\% in the excited state, the intensity of the first peak in the dichroic signal decreases by 14~\%. The shape and intensity of the calculated dichroism are quite sensitive to the crystal-field parameters used in the excited state. 

Nevertheless, despite this slight intensity mismatch with the experimental data, the angular dependence of the crystal is well reproduced by the calculation, which means that the multielectronic approach takes into account the necessary interactions. Since isotropic and dichroic calculated spectra fit well with experiment, the analysis of the calculation is very likely to yield valuable insight into the origin of the experimental transitions in the pre-edge region. In the following, we shall investigate the influence of the different terms in the Hamiltonian taken into account in the LFM  approach (trigonal distortion, fourfold degeneracy of the ground state trigonal S=$\frac{3}{2}$ ($^{4}A_{2g}$), spin-orbit coupling on the 3\textit{d} levels, 3\textit{d}-3\textit{d} or 1\textit{s}-3\textit{d} Coulomb repulsion) on the angular dependence.

\subsubsection{Influence of trigonal distortion on dichroism}
\begin{figure}[!]
\includegraphics[width=7.9cm]{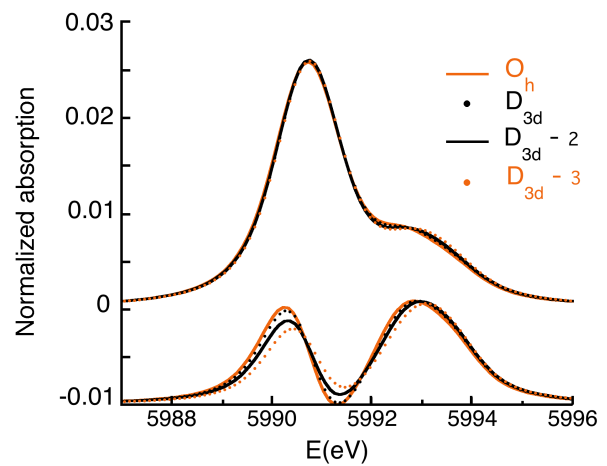}
\caption{\label{fig7} (Color online) Comparison between calculated Cr K pre-edge spectra in spinel in $D_{3d}$ symmetry for increasing trigonal distortion. The orange solid line, labeled $O_h$ corresponds to a perfect \textit{O}$_{h}$ symmetry ($D_{\sigma}$~=~0, $D_{\tau}$ ~=~0). The black dotted line,labeled $D_{3d}$ corresponds to ($D_{\sigma}$~=~-0.036 eV, $D_{\tau}$ ~=~0.089 eV), which are the distortion parameters obtained from Optical Absorption Spectroscopy.\cite{Wood} The black solid line, labeled $D_{3d}$-\textit{2}, was obtained for a doubled distortion parameters ($D_{\sigma}$~=~-0.072 eV, $D_{\tau}$~=~0.178 eV), and the orange dotted line, labeled $D_{3d}$-\textit{3} for tripled distortion parameters ($D_{\sigma}$~=~-0.108 eV, $D_{\tau}$~=~0.267 eV)}
\end{figure}

\
In this paragraph, we investigate the influence of the trigonal distortion on the angular dependence. The isotropic and dichroic spectra in \textit{O}$_{h}$ symmetry have been obtained by setting the trigonal distortion of the crystal field to zero. They are compared to those calculated in $D_{3d}$ symmetry using the distortion parameters given by optical absorption spectroscopy ($D_{\sigma}$ =~-~0.036 eV, $D_{\tau}$~=~0.089 eV). As shown in Fig.~\ref{fig7}, the difference between the calculations performed in $O_h$ and $D_{3d}$ symmetries (orange solid line and black dotted line, respectively) is weak, since the isotropic and dichroic signals have similar shape and intensity. This result is consistent with the small values of the parameters $D_{\sigma}$ and $D_{\tau}$, which quantify the trigonal distortion of the Cr-site in spinel. This means that, provided that trigonal distortion is limited, the calculation of pre-edge spectra could have been performed for a single site with \textit{O}$_{h}$ symmetry (see Appendix A2 for simplified formula). This is also in line with the monoelectronic calculation, for which the splitting between e$^{\downarrow}_{g}$(t$_{2g}$) and a$^{\downarrow}_{1g}$(t$_{2g}$) could not be resolved in the calculated spectra.

We have investigated the effect of the intensity of the trigonal distortion on the calculated spectra by choosing two other sets of the distortion parameters. In Fig.~\ref{fig7}, the spectra labeled $D_{3d}$-\textit{i} (\textit{i} = 2,3) are calculated with the set of distortion parameters (\textit{i}~$\times$~$D_{\sigma}$, \textit{i}~$\times$~$D_{\tau}$), for ($D_{\sigma}$ = - 0.036 eV, $D_{\tau}$ = 0.089 eV). The crystal-field parameters used are the same in the ground- and excited-state. 
As seen in Fig.~\ref{fig7}, the intensity of the isotropic spectra ($D_{3d}$, $D_{3d}$-\textit{2} and $D_{3d}$-\textit{3}) are almost identical, indicating that the isotropic signal is not sensitive to site distortion. The intensity of the maximum at 5990.75~eV remains close to 2.5~\% of the electric dipole edge jump. On the contrary, the shape and intensity of the linear dichroic signal is highly sensitive to the trigonal distortion. The intensity of the first feature at 5990.25~eV is lowered when the distortion is increased: one observes a 20~\% decrease when the distortion parameters are doubled (signal labeled $D_{3d}$-\textit{2} in Fig.\ref{fig7}), and a 50~\% decrease when the distortion parameters are tripled (signal labeled $D_{3d}$-\textit{3}). It should be noticed that, for our parameter sets, the increase of site distortion is accompanied by a rather counter-intuitive decrease of the intensity of the linear dichroic signal, thus indicating the relevance of the theoretical developments performed within this paper. This means that site distortion has to be carefully taken into account when calculations are performed to mimic the angular dependence of the pre-edge. In that case, the calculation for a single site with $D_{3d}$ symmetry should follow the method described in Sec. III.

\subsubsection{Influence of ground-state degeneracy, spin-orbit coupling and interelectronic repulsion on dichroism}\
Beyond the site symmetry distortion, the other ingredients of the calculation are the fourfold degeneracy of the S=$\frac{3}{2}$ ground state, the 3\textit{d}-3\textit{d} Coulomb repulsion, the 1\textit{s}-3\textit{d} Coulomb repulsion and the 3\textit{d} spin-orbit coupling. We shall check the influence of these different parameters. We have performed multiplet calculations restricting the ground-state to the non-degenerate S$_{z}$=$\frac{3}{2}$ state of the S=$\frac{3}{2}$ multiplet. The calculated electric quadrupole transitions are almost identical to those with the four-fold ground state. Differences are below 0.1\% of the maximum intensity of the isotropic electric quadrupole spectrum. This clearly indicates that the procedure followed in monolectronic calculations to take into account the spin degeneracy is sound and appropriate.
The radial integrals for Coulomb interaction and spin-orbit coupling are calculated by relativistic Hartree-Fock atomic calculations. One finds: the 1\textit{s}-3\textit{d} exchange Slater integral $G^{1}_{1s,3d}$~=~0.052 eV, the direct 3\textit{d}-3\textit{d} Slater integrals $F^{2}_{3d,3d}$~=~10.78 eV and $F^{4}_{3d,3d}$~=~6.75 eV, and the 3\textit{d} spin-orbit coupling $\zeta_{3d}$~=~0.035 eV. $G^{1}_{1s,3d}$ is small compared to $F^{2}_{3d,3d}$ and $F^{4}_{3d,3d}$. Using $G^{1}_{1s,3d}$~=~0 in the multiplet calculation, we found almost no difference with the isotropic and dichroic signal calculated with the \textit{ab initio} atomic value of $G^{1}_{1s,3d}$. By calculating the dichroic signal with $\zeta_{3d}$~=~0, we found a small difference concerning the intensity of isotropic and dichroic signals, when compared to the associated spectra with  $\zeta_{3d}$~=~0.035 eV. The maximum relative difference is less than a few percent (2~\%) of the feature intensity. The observed small dependence of the pre-edge features with  $G^{1}_{1s,3d}$ and $\zeta_{3d}$ is in line with results obtained at the Fe K pre-edge. \cite{Arrio} We also performed calculations setting $F^{2}_{3d,3d}$ and $F^{4}_{3d,3d}$ to zero, and we observed that the isotropic and dichroic calculated spectra (not shown) were in complete disagreement with experimental data. This clearly indicates that the direct Slater integrals on the 3\textit{d} shell, and thus the multielectronic 3\textit{d}-3\textit{d} Coulomb interactions, are the essential ingredients governing the shape of the isotropic as well as the dichroic signals.

From the preceding analysis, we can unambiguously determine the parameters governing the shape and intensities of the pre-edge features. Spin-orbit coupling on the 3\textit{d} orbitals, ground-state degeneracy and 1\textit{s}-3\textit{d} Coulomb repulsions have only limited impact on the calculated LFM isotropic spectra. This explains the reasonable agreement between calculation and experiment for isotropic pre-edge in the DFT formulation, where the two first previous ingredients are missing, and where the 1\textit{s}-3\textit{d} Coulomb repulsion is taken into account in an approximate way. The $D_{3d}$ distortion has almost no influence on the isotropic pre-edge but can have a large one on the dichroic signal. In the case of Cr in spinel, the trigonal distortion is such small that it does not provide detectable features on the dichroic signal. The major ingredient for the interpretation of the Cr pre-edge features is 3\textit{d}-3\textit{d} Coulomb repulsion. This effect is highly multielectronic and complicates the simple interpretation provided by the monoelectronic scheme. This ingredient is mandatory to get correct intensities and energies for both isotropic and dichroic signals.  \\

\section{Discussion and conclusion}
On the one hand, monoelectronic calculations allow to make contact between electric dipole and electric quadrupole calculations. They show that electric dipole transitions do not contribute to the features visible in the pre-edge and they provide a clear vision of the assignment of the 1\textit{s}-3\textit{d} transitions occuring in the pre-edge. However, they are unable to reproduce quantitatively the linear dichroism in cubic crystals, since the interelectronic repulsion on the 3\textit{d} levels of the Cr ion cannot fully be described in the LSDA framework. On the other hand, multielectronic calculations well reproduce the angular dependence of the pre-edge in cubic crystals, as well as the isotropic spectrum, with no adjusted parameters. However, in this approach, the main absorption edge, associated to electric dipole transitions, cannot be reproduced since the band structure (or at least the electronic structure of a large enough cluster around the absorbing atom) is not taken into account. The agreement between experiment and multiplet calculations indicates that the assignment of the transitions is no more straightforward, as could have been expected from a more simple atomic monoelectronic picture. Hence, the two approaches are highly complementary. 

From this monoelectronic-multielectronic combined approach, our first finding is that the 3\textit{d}-3\textit{d} electronic repulsions and the crystal field are the main interactions prevailing at the K pre-edge of Cr in spinel. The multiplet approach seems mandatory to describe quantitatively the K pre-edge of 3\textit{d} transition ions, and more generally the K-edge spectra of elements for which electronic correlations are significant. The effect of the 3\textit{d} spin-orbit coupling and of the 1\textit{s}-3\textit{d} Coulomb repulsion are very weak: this explains that the monoelectronic approach (which does not fully take into account these interactions) can provide a satisfactory simulation of the isotropic spectrum. 

Our second finding concerns the maximum proportion of electric quadrupole transitions in the Cr K pre-edge that can be estimated, with respect to the edge jump (here, normalized to 1): the intensity of the largest peak (5990.75~eV) on the pre-edge isotropic spectrum is less than 2.5 \% of the edge jump. From the monoelectronic calculations, we estimate that the non-structured slope from electric dipole origin contributes to about 0.9~\% of the edge jump at 5990.75~eV. Thus, the total intensity of the largest pre-edge feature does not exceed 3.5~\%. We can conclude that, if the pre-edge features are more intense than 4~\% of the edge jump, pure quadrupole transitions alone cannot explain the origin of the structures. It gives a strong limitation to the often encountered idea that electric quadrupole transitions could explain large pre-edge features.
This result, which is consistent with previous studies on Fe$^{2+}$ and Fe$^{3+}$ in minerals and glasses,\cite{Arrio,Westre,Calas} can probably be extended to the other 3\textit{d} transition ions.

Our final finding concerns the relation between the spectral feature of the pre-edge with the local site distortion of the absorbing ion. The effect of the trigonal distortion does not affect significantly the pre-edge isotropic spectrum. This is a general trend already observed for other related spectra (electric dipole transitions for K-edge and L$_{2,3}$ edges of 3\textit{d} elements). On the contrary, the dichroic signal is much more sensitive. This indicates the possibility to obtain quantitative information on site distortion from the linear dichroic dependence of the pre-edge feature. This can only be recorded if angular dependent measurements on single crystals are performed to yield the full dependence of the absorption signal. The connection between site distortion and linear dichroism is then made by simulations in the LFM method within the geometrical analysis developed throughout this paper.
\\

\begin{acknowledgments}
The authors are very grateful to M. Sikora (ID26 beamline) for help during experiment, and to F. Mauri, M. Calandra, M. Lazzeri and C. Gougoussis for fruitful discussions. The theoretical part of this work was supported by the French CNRS computational Institut of Orsay (Institut du D\'eveloppement et de Recherche en Informatique Scientifique) under projects 62015 and 72015. This is IPGP Contribution n$^{\circ}$XXXX. 
\end{acknowledgments}

\newpage

\section{Appendix}
\subsection{Proof of Equation 3}
\subsubsection{General expression of the electric quadrupole cross-section for a cubic crystal}

We start from the defining formula of the electric quadrupole 
cross-section for linearly polarized x-rays:
\begin{eqnarray}
\sigma^{Q}(\bm{\mathrm{\hat{\varepsilon}}},\bm{\mathrm{\hat{k}}}) &=& \pi^2 k^2 \alpha \hbar\omega 
  \sum_f |\langle f |(\bm{\mathrm{\hat{\varepsilon}}}\cdot \mathbf{r} \bm{\mathrm{\hat{k}}} \cdot \mathbf{r}|g\rangle|^2
  \delta(E_g+\hbar\omega-E_f),\nonumber\\
&=& \sum_{ijlm} \varepsilon_i\varepsilon_j k_l k_m \sigma_{iljm},
\label{sigmaijlm}
\end{eqnarray}
where : \\
$\sigma_{ijlm}= \pi^2 k^2 \alpha \hbar\omega  \sum_{f} 
\langle g |r_i r_l|f\rangle\langle f|r_j r_m|g\rangle
  \delta(E_i+\hbar\omega-E_f) $,\\\\
$\bm{\mathrm{\hat{\varepsilon}}}$ and $\bm{\mathrm{\hat{k}}}$ are the polarization and wave unit vectors, respectively. 
 
To calculate the form of this sum when the sample has a symmetry
group $G$, we use the fact that the absorption cross-section is invariant by
any symmetry operation that acts on both the sample variables $\sigma_{ijlm}$
and the x-ray variables $\bm{\mathrm{\hat{\varepsilon}}}, \bm{\mathrm{\hat{k}}}$. Therefore, the cross-section
is left invariant by the crystal symmetries applied to the x-ray variables.
In other words, for any operation $S$ of the symmetry group $G$ of the sample, 
we have $\sigma(\bm{\mathrm{\hat{\varepsilon}}},\bm{\mathrm{\hat{k}}} )=\sigma(S(\bm{\mathrm{\hat{\varepsilon}}}),S(\bm{\mathrm{\hat{k}}}))$.
Therefore, if $G$ is the symmetry group of the sample, or a subgroup of it,
we can write
\begin{eqnarray}
\sigma^{Q}(\bm{\mathrm{\hat{\varepsilon}}},\bm{\mathrm{\hat{k}}}) &=&
\frac{1}{|G|} \sum_S \sigma(S(\bm{\mathrm{\hat{\varepsilon}}}),S(\bm{\mathrm{\hat{k}}})),
\label{sommeG}
\end{eqnarray}
where $|G|$ denotes the number of elements of $G$.\\

We rewrite Eq.~\ref{sigmaijlm} as
\begin{eqnarray}
\sigma^{Q}(\bm{\mathrm{\hat{\varepsilon}}},\bm{\mathrm{\hat{k}}}) & = 
\sum_{i\not= j} \varepsilon_i^2 k_j^2 \sigma_{ijij}
+ \sum_{i\not= j} \varepsilon_i\varepsilon_j k_i k_j \sigma_{iijj}   \nonumber\\
& + \sum_{i\not= j} \varepsilon_i\varepsilon_j k_j k_i \sigma_{ijji}
+ \sum_{i} \varepsilon_i^2 k_i^2 \sigma_{iiii}+ R \nonumber .\\
\label{reste}
\end{eqnarray}
Eq.~\ref{reste} defines term $R$.
The term $R$ is the sum of the terms that
are not ($i=j$ and $l=m$) or ($l=i$ and $m=j$) or ($m=i$ and $l=j$).
So $R$ is a sum of 4 terms of the type
($i\not=j$ and $i\not=l$ and $i\not=m$) plus the three cyclic permutations
of $(i,j,l,m)$ and 4 terms of the type
($i\not=j$ and $j\not=l$ and $l\not=i$) plus the three cyclic permutations
of $(i,j,l,m)$. We want to prove Eq.~\ref{cross_section_quad} that is valid for a cubic crystal in a reference frame such that \textit{x}, \textit{y}, \textit{z} axis are taken along the fourfold symmetry of the cubic crystal. 
We first show that if the sample has three perpendicular mirror planes,
the term $R$ is zero.
Consider the term ($i\not=j$ and $i\not=l$ and $i\not=m$) and take
the symmetry ($S(\varepsilon_i)= -\varepsilon_i$)
the other variables $j,k,l$ are different from $i$, so the symmetry
leaves them invariant and changes only one sign. Therefore, using
Eq.~\ref{sommeG}, this term
disappears from $R$. The same is true for the three cyclic permutations.
Consider now the term
($i\not=j$ and $j\not=l$ and $l\not=i$). Since the values of the indices
is 1, 2 or 3, one of the three indices $i,j,k$ is different from the
other two and from $m$. So one index is again different from the other
ones and the same reasoning can be applied to show that the corresponding
term vanishes. This holds also for the three cyclic permutations and
we have shown that, when there are three perpendicular symmetry planes, the
absorption cross-section is
\begin{eqnarray}
\sigma^Q(\bm{\mathrm{\hat{\varepsilon}}},\bm{\mathrm{\hat{k}}}) = & 
 \sum_{i\not= j} \varepsilon_i^2 k_j^2 \sigma_{ijij}+
 \sum_{i\not= j} \varepsilon_i\varepsilon_j k_i k_j \sigma_{iijj} \nonumber\\ & + 
 \sum_{i\not= j} \varepsilon_i\varepsilon_j k_j k_i \sigma_{ijji}
+ \sum_{i} \varepsilon_i^2 k_i^2 \sigma_{iiii}.
\label{reste2}
\end{eqnarray}
The group $O_h$ has a subgroup made by the six permutations of
$(x,y,z)$.
An average over this subgroup gives the following result
\begin{eqnarray}
\sigma_{cub}^{Q}(\bm{\mathrm{\hat{\varepsilon}}},\bm{\mathrm{\hat{k}}}) = & 
 \sum_{i\not= j} \varepsilon_i^2 k_j^2 A+
 \sum_{i\not= j} \varepsilon_i\varepsilon_j k_i k_j B \nonumber\\
 & +  \sum_{i} \varepsilon_i^2 k_i^2 C,
\label{reste3}
\end{eqnarray}
where
\begin{eqnarray*}
A &=& \sum_{i\not= j} \frac{\sigma_{ijij}}{6},\\
B &=& \sum_{i\not= j} \frac{\sigma_{iijj}+\sigma_{ijji}}{3},\\
C &=& \sum_{i} \frac{\sigma_{iiii}}{3}.
\end{eqnarray*}
To complete the proof, we use the fact that
\begin{eqnarray*}
 \sum_{i\not= j} \varepsilon_i^2 k_j^2 &=& \bm{\mathrm{\hat{\varepsilon}}}\cdot\bm{\mathrm{\hat{\varepsilon}}} \bm{\mathrm{\hat{k}}}\cdot\bm{\mathrm{\hat{k}}}
- \sum_{i} \varepsilon_i^2 k_i^2,\\
 \sum_{i\not= j} \varepsilon_i\varepsilon_j k_i k_j &=&
  (\bm{\mathrm{\hat{\varepsilon}}}\cdot \bm{\mathrm{\hat{k}}})^2 -\sum_{i} \varepsilon_i^2 k_i^2,
\end{eqnarray*}
and the identities
$\bm{\mathrm{\hat{\varepsilon}}}\cdot\bm{\mathrm{\hat{\varepsilon}}}=1$, $\bm{\mathrm{\hat{k}}}\cdot\bm{\mathrm{\hat{k}}}=1$ and
$\bm{\mathrm{\hat{\varepsilon}}}\cdot\bm{\mathrm{\hat{k}}}=0$ to get
\begin{eqnarray}
\sigma_{cub}^{Q}(\bm{\mathrm{\hat{\varepsilon}}},\bm{\mathrm{\hat{k}}}) &=&  A + \sum_{i} \varepsilon_i^2 k_i^2 (C-A-B).
\label{reste4}
\end{eqnarray}

To compare this result with the expansion over spherical tensors,
we need to determine the isotropic contribution 
$\sigma^Q_0$, which is obtained as the average
of $\sigma^Q_{cub}(\bm{\mathrm{\hat{\varepsilon}}},\bm{\mathrm{\hat{k}}})$ over angles.
We write  $\sigma(\bm{\mathrm{\hat{\varepsilon}}},\bm{\mathrm{\hat{k}}})$ in terms of $\theta$, $\phi$, $\psi$ as in Sec. III A 2.

Thus,
\begin{eqnarray*}
& \sigma_{cub}^Q(\bm{\mathrm{\hat{\varepsilon}}},\bm{\mathrm{\hat{k}}}) =
A + \frac{C-A-B}{4} \sin^2\theta
 \big(7\cos^2\theta\cos^2\psi \\
 \nonumber & +\cos 4\phi \cos^2\theta\cos^2\psi) + 2\sin^2 2\phi \sin^2\psi - \cos\theta \sin 4\phi \sin 2\psi\big),
\end{eqnarray*}
and the average over all directions is 
\begin{eqnarray*}
\sigma^Q_0 &= \frac{1}{8\pi^2}
  \int_0^\pi \sin\theta d\theta 
  \int_0^{2\pi} d\phi
  \int_0^{2\pi} d\psi \sigma^Q_{cub}(\bm{\mathrm{\hat{\varepsilon}}},\bm{\mathrm{\hat{k}}})\\
 \nonumber & = A + \frac{C-A-B}{5}.
\end{eqnarray*}
Therefore,
\begin{eqnarray*}
\sigma_{cub}^Q(\bm{\mathrm{\hat{\varepsilon}}},\bm{\mathrm{\hat{k}}}) &=& 
\sigma^Q_0 + \big(\sum_{i} \varepsilon_i^2 k_i^2 -\frac{1}{5}\big)(C-A-B)
\end{eqnarray*}

and 

\begin{equation}
\label{preuve}
\sigma_{cub}^{Q}(\hat{\bm{\varepsilon}},\hat{\textbf{k}})= \sigma^{Q}_0 +(\varepsilon_{x}^{2} k_{x}^{2} +\varepsilon_{y}^{2} k_{y}^{2} + \varepsilon_{z}^{2} k_{z}^{2}-\frac{1}{5} )\sigma_{1}^{Q},
\end{equation}
where $\sigma_1^Q$ = $(C - A - B)$ and $k_x$, $k_y$, $k_z$ are the coordinates of $\bm{\mathrm{\hat{k}}}$.\\

This expression is valid for any cubic crystal, providing that the reference frame is such that the \textit{x}, \textit{y}, \textit{z} axis are taken along the fourfold symmetry axis of the cubic crystal.

\subsubsection {Absorption cross-sections in $O_{h}$ symmetry\\}
For a site with $O_{h}$ symmetry, the orthonormal reference frame chosen is that of the cubic crystal. The \textit{z}-axis of the reference frame is parallel to the fourfold axis of the cube. The angle $\theta$ is thus the angle between the polarization vector and the \textit{z}-axis of the cube.

The cross-section calculated for a single site in $O_{h}$ is equal to the cross-section of the cubic crystal, since a perfect octahedron and the cube have the same symmetry operations: \\

$\sigma_{O_{h}}^{Q}$($\hat{\bm{\varepsilon}}$,$\hat{\textbf{k}}$) =~$\sigma_{cub}^{Q}$($\hat{\bm{\varepsilon}}$,$\hat{\textbf{k}}$) and 
$\sigma_{O_{h}}^{D}$($\hat{\bm{\varepsilon}}$) =~$\sigma_{cub}^{D}$($\hat{\bm{\varepsilon}}$).
\\

For a single site with \textit{O}$_{h}$ symmetry, the expression of the electric dipole cross-section is very simple: 
\begin{equation}
\sigma_{O_{h}}^{D} (\hat{\bm{\varepsilon}}) = \sigma^{D}(0,0), 
\end{equation}
where $\sigma^{D}(0,0)$ is the isotropic electric dipole cross-section.\\

For the cubic crystal, one obtains:
\begin{equation}
\sigma_{cub}^{D} (\hat{\bm{\varepsilon}}) = \sigma_{O_{h}}^{D}(\hat{\bm{\varepsilon}}) = \sigma^{D}(0,0).
\end{equation}

The electric quadrupole absorption cross-section for a site with $O_{h}$ symmetry is given by: 
\begin{align}
\label{cross_section_Oh}
 \nonumber\sigma_{O_{h}}^{Q}(\hat{\bm{\varepsilon}},\hat{\textbf{k}})&=
 \sigma_{O_{h}}^{Q}(0,0)+\frac{1}{\sqrt{14}}~[35~\sin^{2}\theta~\cos^{2}\theta~\cos^{2}\psi  \\
\nonumber &+ 5~\sin^{2}\theta~\sin^{2}\psi-4\\
 \nonumber &+5~\sin^{2}\theta(\cos^{2}\theta~\cos^{2}\psi~\cos4~\phi-\sin^{2}\psi~\cos4\phi)\\
& -2~\cos\theta~\sin\psi~\cos\psi ~\sin4\phi~]~\sigma_{O_{h}}^{Q}(4,0),
\end{align}
where $\sigma_{cub}^{Q}(0,0)$ is the isotropic electric quadrupole cross-section, and $\sigma_{cub}^{Q}(4,0)$ a purely anisotropic electric quadrupole term.\\

Using $\sigma_{O_{h}}^{Q}$($\hat{\bm{\varepsilon}}$,$\hat{\textbf{k}}$) =~$\sigma_{cub}^{Q}$($\hat{\bm{\varepsilon}}$,$\hat{\textbf{k}}$) and ($\theta=\frac{\pi}{2}$, $\phi=\frac{\pi}{2}$, $\psi=\frac{\pi}{2}$), Eq.~\ref{cross_section_Oh} is equivalent to Eq.~\ref{preuve} with:
\begin{align}
\label{Oh}&\sigma_{0}^{Q} = \sigma_{O_{h}}^{Q}(0,0) = \sigma_{cub}^{Q}(0,0), \\
\label{Oh2}&\sigma_{1}^{Q} =~\frac{20}{\sqrt{14}}~\sigma_{O_{h}}^{Q}(4,0) = ~\frac{20}{\sqrt{14}}~\sigma_{cub}^{Q}(4,0).
\end{align}\\

Eq.~\ref{preuve} can be rewritten as: 
\begin{align}
\nonumber \sigma_{cub}^{Q}(\hat{\bm{\varepsilon}},\hat{\textbf{k}}) &=  \sigma_{cub}^{Q}(0,0)\\
& +\frac{20}{\sqrt{14}}(\varepsilon_{x}^{2} k_{x}^{2} +\varepsilon_{y}^{2} k_{y}^{2} + \varepsilon_{z}^{2} k_{z}^{2}-\frac{1}{5} )\sigma_{cub}^{Q}(4,0).
\end{align}  

This is the proof of Eq.~\ref{cross_section_quad}.

\subsection {Expression of $\bm{\mathrm{\hat{\varepsilon}}}$($\alpha_{rot}$) and $\bm{\mathrm{\hat{k}}}$ ($\alpha_{rot}$)\\}

For the sample cut and the experimental setup used in this study, we have: 
\begin{eqnarray*}
\bm{\mathrm{\hat{\varepsilon}}}(\alpha_{rot}) = \begin{pmatrix}
\frac{1-\cos\alpha_{rot}}{2} \\
\frac{1+\cos\alpha_{rot}}{2} \\
\frac{\sin\alpha_{rot}}{\sqrt{2}} 
\end{pmatrix},\\
\bm{\mathrm{\hat{k}}} (\alpha_{rot}) = \begin{pmatrix}
\frac{-1-\cos\alpha_{rot}}{2} \\
\frac{-1+\cos\alpha_{rot}}{2} \\
\frac{\sin\alpha_{rot}}{\sqrt{2}}
\end{pmatrix}.
\end{eqnarray*}\\

\subsection{Symmetry adapted method used in monoelectronic calculations}

Firstly, the absorption cross-section was calculated in the electric dipole approximation, in order to derive $\sigma$$_{cub}^{D}(0,0)$, the isotropic electric dipole cross-section. This term can be obtained from a single calculation of the electric dipole absorption cross-section performed for site A. The expression of the electric dipole cross-section in $D_{3d}$, $\sigma_{A}^{D}$, is given by Eq.~\ref{dipD3d}, where $\theta$ is the angle between the polarization vector and the $C_{3}$ axis. 
The program we used for the \textit{ab initio} calculations calculates the average of $\sigma_{A}^{D}$ ($\bm{\mathrm{\hat{\varepsilon}}}$) for $\bm{\mathrm{\hat{\varepsilon}}}$ along the \textit{x}-, \textit{y}- and \textit{z}-axis of the cubic frame. The angle $\theta$ between the [$\bar{1}$11] direction (parallel to the $C_{3}$ axis of the site) and each of these three directions, is $\arccos \frac{1}{\sqrt{3}}$. This implies that, for a polarization vector $\bm{\mathrm{\hat{\varepsilon}}}$ taken along the \textit{x}-, \textit{y}- or \textit{z}-axis,
\begin{equation}
\sigma_{cub}^{D} (\bm{\mathrm{\hat{\varepsilon}}}) = \sigma_{A}^{D} (\bm{\mathrm{\hat{\varepsilon}}}) = \sigma_{D_{3d}}^{D}(0,0) = \sigma_{cub}^{D}(0,0).
\end{equation}
This means that the average value calculated by the program is directly equal to $\sigma_{cub}^{D}(0,0)$. Hence, $\sigma_{cub}^{D}(0,0)$ was obtained from a single calculation performed at site A. 

Secondly, the calculation was performed in the electric quadrupole approximation, in order to derive $\sigma$$_{cub}^{Q}(0,0)$ and $\sigma$$_{cub}^{Q}(4,0)$. Once these two terms are determined, we will be able to calculate the electric quadrupole absorption cross-section for the cubic crystal, according to Eq. \ref{cross_section_quad}, for any ($\bm{\mathrm{\hat{\varepsilon}}}$,$\bm{\mathrm{\hat{k}}}$) configuration. We used a symmetry adapted method to determine $\sigma$$_{cub}^{Q}(0,0)$ and $\sigma$$_{cub}^{Q}(4,0)$ in order to reduce the number of calculations: this way, it is possible to consider one single substitutional site (site A) and take advantage of the symmetry properties of the crystal. This method allows to save significant computational time, since we perform only two self-consistent calculations (instead of eight with the \textit{brute force} method): one calculation to do the structural relaxation of the system (substituted at site A), and a second one to calculate the charge density with a core-hole on Cr.
As mentioned in Ref.~\onlinecite{Brouder2}, assuming that we have calculated the spectrum for a given site X, it is possible to obtain the spectrum for any site Y equivalent to X by calculating the spectrum of site X for a rotated x-ray beam. More precisely, in the case of electric quadrupole transitions, if site Y is the image of site X by a rotation R, the spectrum of site Y for a configuration ($\bm{\mathrm{\hat{\varepsilon}}}$,$\bm{\mathrm{\hat{k}}}$) is equal to the spectrum of site X for the rotated configuration (R$^{-1}$($\bm{\mathrm{\hat{\varepsilon}}}$),R$^{-1}$($\bm{\mathrm{\hat{k}}}$)). As in Ref. \onlinecite{Brouder2}, let us consider the site with reduced coordinates (0, 1/4, 3/4), with is a representative of site A. By applying the three rotations about the \textit{z}-axis of the crystal through angles of $\pi$/2, $\pi$, 3$\pi$/2, we obtain the positions of three sites, which are respectively representative of sites B, D and C. The three rotations will denoted as R$_{\pi/2}$, R$_{\pi}$ and R$_{3\pi/2}$. 
This implies that 
\begin{eqnarray}
\sigma_{B}^{Q}(\bm{\mathrm{\hat{\varepsilon}}},\bm{\mathrm{\hat{k}}})&=& \sigma_{A}^{Q}(R^{-1}_{\pi/2}(\bm{\mathrm{\hat{\varepsilon}}}),R^{-1}_{\pi/2}(\bm{\mathrm{\hat{k}}})),\\
\sigma_{C}^{Q}(\bm{\mathrm{\hat{\varepsilon}}},\bm{\mathrm{\hat{k}}})&=& \sigma_{A}^{Q}(R^{-1}_{3\pi/2}(\bm{\mathrm{\hat{\varepsilon}}}),R^{-1}_{3\pi/2}(\bm{\mathrm{\hat{k}}})),\\
\sigma_{D}^{Q}(\bm{\mathrm{\hat{\varepsilon}}},\bm{\mathrm{\hat{k}}})&=& \sigma_{A}^{Q}(R^{-1}_{\pi}(\bm{\mathrm{\hat{\varepsilon}}}),R^{-1}_{\pi}(\bm{\mathrm{\hat{k}}})).
\end{eqnarray}

We shall now apply the previous equations to the case of the two experimental configurations,  $\alpha_{rot}$~=~0$^{\circ}$ and $\alpha_{rot}$~=~90$^{\circ}$, in order to see if additional simplifications can be found. To do so, we need to consider the expression of the electric quadrupole transition operator \textit{$\mathrm{\hat{Q}}$}: \\
\begin{equation}
\hat{Q} = \frac{i}{2}\bm{\mathrm{\hat{\varepsilon}}}\cdot\textbf{r}\bm{\mathrm{\hat{k}}}\cdot\textbf{r}.
\end{equation} 
Note that, in text, the assignment of the calculated monoelectronic transitions is discussed using $\hat{O}=\bm{\mathrm{\hat{\varepsilon}}}\cdot\textbf{r}\bm{\mathrm{\hat{k}}}\cdot\textbf{r}$. We have to consider the absolute value of $\hat{O}$ (or $\hat{Q}$) because the cross-section does not depend on the sign of $\hat{O}$.\\\\
For $\alpha_{rot}$~=~0$^{\circ}$ ($\bm{\mathrm{\hat{\varepsilon}}}$~=~[010], $\bm{\mathrm{\hat{k}}}$~=~[$\bar{1}$00]): 

\begin{table*}[!t]
\caption{Expression of the electric quadrupole cross-section calculated in $D_{3d}$ group for the four independent orientations used to derive $\sigma_{D_{3d}}^{Q}(0,0)$, $\sigma_{D_{3d}}^{Q}(2,0)$, $\sigma_{D_{3d}}^{Q}(4,0)$ and $\sigma_{D_{3d}}^{Q}(4,3)$ (see text). The coordinates of $\bm{\mathrm{\hat{\varepsilon}}}$ and $\bm{\mathrm{\hat{k}}}$ are given in the reference frame chosen for $D_{3d}$, with the \textit{z}-direction parallel to the $C_{3}$ axis.}
\label{tab:D3d}
\begin{ruledtabular}
\begin{tabular}{ccccccc}
label &$\theta$ & $\phi$ & $\psi$ & $\bm{\mathrm{\hat{\varepsilon}}}$ & $\bm{\mathrm{\hat{k}}}$ & $\sigma_{D_{3d}}^{Q}(\bm{\mathrm{\hat{\varepsilon}}},\bm{\mathrm{\hat{k}}})$ \\ \hline\\
\textit{s1} & $\arccos(\frac{1}{\sqrt{3}})$ & $-\frac{2\pi}{3}$ & $\frac{\pi}{4}$ & (-$\frac{1}{\sqrt{6}}$, -$\frac{1}{\sqrt{2}}$,$\frac{1}{\sqrt{3}}$)& ($\frac{1}{\sqrt{6}}$,-$\frac{1}{\sqrt{2}}$,-$\frac{1}{\sqrt{3}}$)  & $\sigma_{D_{3d}}^{Q}$(0,0)+$\frac{\sqrt{14}}{9}$$\sigma_{D_{3d}}^{Q}(4,0)$+ $\frac{4\sqrt{5}}{9}$$\sigma_{D_{3d}}^{Q}$(4,3)\\
\textit{s2} & $\arccos(\frac{1}{\sqrt{3}}+\frac{1}{\sqrt{6}})$& 0 & 0  &  ($\frac{-1+\sqrt{2}}{\sqrt{6}}$,0,$\frac{1+\sqrt{2}}{\sqrt{6}}$) &($\frac{1+\sqrt{2}}{\sqrt{6}}$,0,$\frac{1-\sqrt{2}}{\sqrt{6}}$)&  $\sigma_{D_{3d}}^{Q}$(0,0)-$\sqrt{\frac{5}{14}}$$\sigma_{D_{3d}}^{Q}(2,0)$- $\frac{109}{36\sqrt{14}}$$\sigma_{D_{3d}}^{Q}$(4,0)-$\frac{2\sqrt{5}}{9}$$\sigma_{D_{3d}}^{Q}$(4,3)\\
\textit{s3} & $\frac{\pi}{2}$ & $\frac{\pi}{2}$ &$\frac{\pi}{2}$  &  (0,1,0) &(-1,0,0)&  $\sigma_{D_{3d}}^{Q}$(0,0)+$\sqrt{\frac{10}{7}}$$\sigma_{D_{3d}}^{Q}(2,0)$+ $\frac{1}{\sqrt{14}}$$\sigma_{D_{3d}}^{Q}$(4,0)\\
\textit{s4} & $\frac{3\pi}{4}$ & $\frac{\pi}{2}$ &$\pi$  &  (0,$\frac{1}{\sqrt{2}}$,-$\frac{1}{\sqrt{2}}$) &(0,$\frac{1}{\sqrt{2}}$,$\frac{1}{\sqrt{2}}$)&  $\sigma_{D_{3d}}^{Q}$(0,0)-$\sqrt{\frac{5}{14}}$$\sigma_{D_{3d}}^{Q}(2,0)$+ $\frac{19}{4\sqrt{14}}$$\sigma_{D_{3d}}^{Q}$(4,0)\\
\end{tabular}
\end{ruledtabular}
\end{table*}

\begin{align}
\nonumber 
|\bm{\mathrm{\hat{\varepsilon}}}\cdot\textbf{r}\bm{\mathrm{\hat{k}}}\cdot\textbf{r}|&=|R^{-1}_{\pi/2}(\bm{\mathrm{\hat{\varepsilon}}})\cdot\textbf{r}R^{-1}_{\pi/2}(\bm{\mathrm{\hat{k}}})\cdot\textbf{r}|=|R^{-1}_{\pi}(\bm{\mathrm{\hat{\varepsilon}}})\cdot\textbf{r}R^{-1}_{\pi}(\bm{\mathrm{\hat{k}}})\cdot\textbf{r}| \\
\nonumber &= |R^{-1}_{3\pi/2}(\bm{\mathrm{\hat{\varepsilon}}})\cdot\textbf{r}R^{-1}_{\pi/2}(\bm{\mathrm{\hat{k}}})\cdot\textbf{r}| = xy. 
\end{align}
Hence, we have: 
\begin{align} 
\nonumber \sigma_{A}^{Q}(\bm{\mathrm{\hat{\varepsilon}}},\bm{\mathrm{\hat{k}}})& = \sigma_{A}^{Q}(R^{-1}_{\pi/2}(\bm{\mathrm{\hat{\varepsilon}}}),R^{-1}_{\pi/2}(\bm{\mathrm{\hat{k}}}))\\
\nonumber & =\sigma_{A}^{Q}(R^{-1}_{\pi}(\bm{\mathrm{\hat{\varepsilon}}}),R^{-1}_{\pi}(\bm{\mathrm{\hat{k}}}))\\
\nonumber &=\sigma_{A}^{Q}(R^{-1}_{3\pi/2}(\bm{\mathrm{\hat{\varepsilon}}}),R^{-1}_{3\pi/2}(\bm{\mathrm{\hat{k}}})).
\end{align}
This means that: \\
\begin{equation}
\sigma_{A}^{Q}(\bm{\mathrm{\hat{\varepsilon}}},\bm{\mathrm{\hat{k}}}) = \sigma_{B}^{Q}(\bm{\mathrm{\hat{\varepsilon}}},\bm{\mathrm{\hat{k}}}) = \sigma_{C}^{Q}(\bm{\mathrm{\hat{\varepsilon}}},\bm{\mathrm{\hat{k}}}) =\sigma_{D}^{Q}(\bm{\mathrm{\hat{\varepsilon}}},\bm{\mathrm{\hat{k}}}).
\end{equation}
For $\alpha_{rot}$~=~0$^{\circ}$, we thus need to perform one single calculation of the electric quadrupole cross-section, 
since 
\begin{equation}
\sigma_{cub}^{Q}(\bm{\mathrm{\hat{\varepsilon}}},\bm{\mathrm{\hat{k}}})= \sigma_{A}^{Q}(\bm{\mathrm{\hat{\varepsilon}}},\bm{\mathrm{\hat{k}}}).
\end{equation}\\
For $\alpha_{rot}$~=~90$^{\circ}$ ($\bm{\mathrm{\hat{\varepsilon}}}$ = [$\frac{1}{2}$,$\frac{1}{2}$,$\frac{1}{\sqrt{2}}$]  and $\bm{\mathrm{\hat{k}}}$ = [-$\frac{1}{2}$,-$\frac{1}{2}$,$\frac{1}{\sqrt{2}}$]): 

\begin{align}
\nonumber & |\bm{\mathrm{\hat{\varepsilon}}}\cdot\textbf{r}\bm{\mathrm{\hat{k}}}\cdot\textbf{r}| =  |R^{-1}_{\pi}(\bm{\mathrm{\hat{\varepsilon}}})\cdot\textbf{r}R^{-1}_{\pi}(\bm{\mathrm{\hat{k}}})\cdot\textbf{r}| = |z^{2}/2-(x-y)^{2}/4|  \\
\nonumber & \mathrm{and} \\
\nonumber & |R^{-1}_{\pi/2}(\bm{\mathrm{\hat{\varepsilon}}})\cdot\textbf{r}R^{-1}_{\pi/2}(\bm{\mathrm{\hat{k}}})\cdot\textbf{r}| = |R^{-1}_{3\pi/2}(\bm{\mathrm{\hat{\varepsilon}}})\cdot\textbf{r}R^{-1}_{\pi/2}(\bm{\mathrm{\hat{k}}})\cdot\textbf{r}| \\
\nonumber &= |z^{2}/2-(x+y)^{2}/4|.
\end{align}

This means that: 
\begin{eqnarray}
\sigma_{A}^{Q}(\bm{\mathrm{\hat{\varepsilon}}},\bm{\mathrm{\hat{k}}}) =\sigma_{D}^{Q}(\bm{\mathrm{\hat{\varepsilon}}},\bm{\mathrm{\hat{k}}}),\\
\sigma_{B}^{Q}(\bm{\mathrm{\hat{\varepsilon}}},\bm{\mathrm{\hat{k}}}) = \sigma_{C}^{Q}(\bm{\mathrm{\hat{\varepsilon}}},\bm{\mathrm{\hat{k}}}). 
\end{eqnarray}
For $\alpha_{rot}$~=~90$^{\circ}$, we thus need to perform two calculations of the electric quadrupole cross-sections  
since 
\begin{equation}
\sigma_{cub}^{Q}(\bm{\mathrm{\hat{\varepsilon}}},\bm{\mathrm{\hat{k}}})= \frac{\sigma_{A}^{Q}(\bm{\mathrm{\hat{\varepsilon}}},\bm{\mathrm{\hat{k}}}) + \sigma_{C}^{Q}(\bm{\mathrm{\hat{\varepsilon}}},\bm{\mathrm{\hat{k}}})}{2} 
\end{equation}
As mentioned previously, instead of doing the calculation for the two sites A and C, it is more convenient to compute the spectrum for site A, for the two orientations ($\bm{\mathrm{\hat{\varepsilon}}}$,$\bm{\mathrm{\hat{k}}}$) and (R$^{-1}_{\frac{-3\pi}{2}}$($\bm{\mathrm{\hat{\varepsilon}}}$),R$^{-1}_{\frac{-3\pi}{2}}$($\bm{\mathrm{\hat{k}}}$)). This corresponds to ($\bm{\mathrm{\hat{\varepsilon}}}$ = [$\frac{1}{2}$,$\frac{1}{2}$,$\frac{1}{\sqrt{2}}$], $\bm{\mathrm{\hat{k}}}$ = [-$\frac{1}{2}$,-$\frac{1}{2}$,$\frac{1}{\sqrt{2}}$]) and ($\bm{\mathrm{\hat{\varepsilon}}}$ = [-$\frac{1}{2}$,$\frac{1}{2}$,$\frac{1}{\sqrt{2}}$], $\bm{\mathrm{\hat{k}}}$ = [$\frac{1}{2}$,-$\frac{1}{2}$,$\frac{1}{\sqrt{2}}$]), respectively.\\\\

\subsection{Method used to perform the multiplet calculations}

As mentioned in Sec. III A 2 (Eq.~\ref{cross_section_D3d}), one needs first to determine $\sigma_{D_{3d}}^{Q}(0,0)$, $\sigma_{D_{3d}}^{Q}(2,0)$, $\sigma_{D_{3d}}^{Q}(4,0)$ and $\sigma_{D_{3d}}^{Q}(4,3)$, in order to determine $\sigma_{cub}^{Q}(0,0)$ and $\sigma_{cub}^{Q}(4,0)$  using Eqs~\ref{moyenne1}~and~\ref{moyenne2}. This is done by performing four multiplet calculations, which provide four independent values of the electric quadrupole cross-section, \textit{s1}, \textit{s2}, \textit{s3} and \textit{s4}, where $\bm{\mathrm{\hat{\varepsilon}}}$ and $\bm{\mathrm{\hat{k}}}$ are defined in Table ~\ref{tab:D3d}. The components $\sigma_{D_{3d}}^{Q}(0,0)$, $\sigma_{D_{3d}}^{Q}(2,0)$, $\sigma_{D_{3d}}^{Q}(4,0)$ and $\sigma_{D_{3d}}^{Q}(4,3)$ are obtained by a combination of \textit{s1}, \textit{s2}, \textit{s3} and \textit{s4}, according to: 
\begin{align}
\nonumber&\sigma_{D_{3d}}^{Q}(0,0) =\frac{1}{5}~\textit{s1}+\frac{2}{5}~ \textit{s2}+\frac{4}{15}~\textit{s3}+\frac{2}{15}~\textit{s4},\\
\nonumber&\sigma_{D_{3d}}^{Q}(2,0) =-\frac{1}{\sqrt{70}}~\textit{s1}-\sqrt{\frac{2}{35}}~ \textit{s2}+\sqrt{\frac{5}{14}}~\textit{s3}-\sqrt{ \frac{2}{35}}~\textit{s4},\\
\nonumber&\sigma_{D_{3d}}^{Q}(4,0) =-\frac{2\sqrt{14}}{35}~\textit{s1}-\frac{4\sqrt{14}}{35}~ \textit{s2}+\frac{2\sqrt{14}}{105}~\textit{s3}+\frac{16\sqrt{14}}{105}~\textit{s4},\\
\nonumber&\sigma_{D_{3d}}^{Q}(4,3) =\frac{2}{\sqrt{5}}~\textit{s1}-\frac{1}{2\sqrt{5}}~ \textit{s2}-\frac{2}{3\sqrt{5}}~\textit{s3}-\frac{\sqrt{5}}{6}~\textit{s4}.\\
\end{align}
These equations have been obtained by inversing the system of equations, which give the expressions of \textit{s1}, \textit{s2}, \textit{s3} and \textit{s4}, in function of $\sigma_{D_{3d}}^{Q}(0,0)$, $\sigma_{D_{3d}}^{Q}(2,0)$, $\sigma_{D_{3d}}^{Q}(4,0)$ and $\sigma_{D_{3d}}^{Q}(4,3)$ (Table ~\ref{tab:D3d}).

Once this first step has been performed, Eq. \ref{moyenne1} and \ref{moyenne2} are used to derive $\sigma_{cub}^{Q}(0,0)$ and $\sigma_{cub}^{Q}(4,0)$. The electric quadrupole cross-section of the cubic crystal can then be calculated for any experimental configuration using Eq.~\ref{alpha}. 

\subsection{Expression of the \textit{d}-eigenstates in $D_{3d}$}

The \textit{d}-eigenstates in $D_{3d}$ point group are determined by the branching rules of the irreducible representation 2$^{+}$(O$_{3}$) in $D_{3d}$. In order to get the complete eigenstates, we must consider the $O_3\supset O_h \supset D_{3d} \supset C_{3i}$ subduction. To simplify the notation, we will make no use of the parity ($\pm$ or $g/u$) in the rest of the appendix. Hence, we will use the subduction $SO_3\supset O \supset D_{3} \supset C_{3}$. In the following, the irreducible representations are labeled according to Ref.~\onlinecite{Butler}. For example, in $O$ group, the irreducible representation $\tilde{1}$ designates $T_2$ in Sch{\"o}nflies notation, while 2 designates E. The complete eigenstates are written as $|k(SO_3)\rho(O)\sigma(D_{3})\lambda(C_{3})\rangle_3$, where $\lambda$ is the irreducible representation of $C_{3}$ subgroup, coming from the \textit{k} irreducible representation of $SO_{3}$, that becomes $\rho$ in \textit{O}, $\sigma$ in $D_{3}$ and $\lambda$ in $C_{3}$. The branching rules for $k$= 2 are given in Tab.~\ref{k=2}.

\begin{table}[!t]
\caption{Branching rules for $k$ = 2 using the $SO_3\supset O \supset D_{3} \supset C_{3}$ subduction}
\label{k=2}
$$
\begin{array}{|c @{\quad \rightarrow \quad} 
               c @{\quad \rightarrow \quad}   c @{\quad \rightarrow \quad}
                              c @{\quad} |c|} 
\hline
  SO_3 & O & D_3 &  C_3  
& |k(SO_3)\rho(O)\sigma(D_3)\lambda(C_3)\rangle\\
\hline
2 & \tilde{1}    &    1    &  1 &  |e_+(t_2)\rangle\\
2 & \tilde{1}    &    1     &  -1 & |e_{-}(t_{2})\rangle\\
2 & \tilde{1}  &    0   &  0 & |a_{1}(t_{2})\rangle\\ 
2 &    2            &    1    & 1 & |e_{+}(e)\rangle\\
2 &    2            &    1    & -1 & |e_{-}(e)\rangle\\
 \hline
\end{array}
$$
\end{table}

We recall that in $D_{3}$ (or $C_{3}$), the reference frame chosen is not the one used in $O$. Therefore, we need to express the $|k(SO_3)\rho(O)\sigma(D_{3})\lambda(C_{3})\rangle$$_{3}$ in function of $|k(SO_3)\rho(O)\sigma(D_{4})\lambda(C_{4})\rangle$$_{4}$, where $|k(SO_3)\rho(O)\sigma(D_{4})\lambda(C_{4})\rangle$$_{4}$ are determined using the $SO_3\supset O \supset D_{4} \supset C_{4}$ subduction. To do so, we use the relations given for \textit{k}~=~2 in Ref.~\onlinecite{Butler} (p. 549): 
\begin{align}
\label{C3states}
\nonumber  &{|2211\rangle}_3 = -\frac{1}{\sqrt{2}} {|2200\rangle}_4  + i \frac{1}{\sqrt{2}} {|2222\rangle}_4\\
\nonumber  &{|221\!-\!1\rangle}_3 = -\frac{1}{\sqrt{2}} {|2200\rangle}_4  - i \frac{1}{\sqrt{2}} {|2222\rangle}_4\\
\nonumber  &{|2\tilde{1}00\rangle}_3 = (1+i) \frac{1}{\sqrt{6}} {|2\tilde{1}11\rangle}_4  \\
\nonumber &~~~~~~~~~ + (1-i) \frac{1}{\sqrt{6}} {|2\tilde{1}1\!-\!1\rangle}_4 -\frac{i}{\sqrt{3}}{|2\tilde{1}\tilde{2}2\rangle}_4\\
\nonumber  &{|2\tilde{1}11\rangle}_3 = (1+i) (\frac{1}{2\sqrt{2}}-\frac{1}{2\sqrt{6}}) {|2\tilde{1}11\rangle}_4  \\
\nonumber  &~~~~~~~~~+ (-1+i) (\frac{1}{2\sqrt{2}}+\frac{1}{2\sqrt{6}}) {|2\tilde{1}1\!-\!1\rangle}_4 -\frac{i}{\sqrt{3}} {|2\tilde{1}\tilde{2}2\rangle}_4\\
\nonumber  &{|2\tilde{1}1\!-\!1\rangle}_3 = -(1+i) (\frac{1}{2\sqrt{2}}+\frac{1}{2\sqrt{6}}) {|2\tilde{1}11\rangle}_4 \\ 
\nonumber  &~~~~~~~~~+ (1-i) (\frac{1}{2\sqrt{2}}-\frac{1}{2\sqrt{6}}) {|2\tilde{1}1\!-\!1\rangle}_4 -\frac{i}{\sqrt{3}} {|2\tilde{1}\tilde{2}2\rangle}_4\\
\end{align}

If we now express the $|k(SO_3)\rho(O)\sigma(D_{4})\lambda(C_{4})\rangle$$_{4}$ as $|JM\rangle$ partners (Ref.~\onlinecite{Butler} p. 527), we obtain:
\begin{align}
\label{Ostates}
\nonumber & {|2200\rangle}_4  = - |20\rangle\\
\nonumber & {|2222\rangle}_4 = - \frac{1}{\sqrt{2}} |22\rangle - \frac{1}{\sqrt{2}} |2\!-\!2\rangle\\
\nonumber & {|2\tilde{1}11\rangle}_4  = - |21\rangle\\
\nonumber & {|2\tilde{1}1\!-\!1\rangle}_4  =  |2\!-\!1\rangle\\
\nonumber & {|2\tilde{1}\tilde{2}2\rangle}_4 =  \frac{1}{\sqrt{2}} |22\rangle - \frac{1}{\sqrt{2}} |2\!-\!2\rangle\\
\end{align}

In $O$, the \textit{d} orbitals are expressed as: 
\begin{align}
\label{orbd}
\nonumber & d_{xy} =  \frac{i}{\sqrt{2}} (|2\!-\!2\rangle-|22\rangle)\\
\nonumber & d_{yz}  = \frac{i}{\sqrt{2}} (|2\!-\!1\rangle+|21\rangle)\\
\nonumber & d_{xz}  = \frac{1}{\sqrt{2}} (|2\!-\!1\rangle-|21\rangle)\\
\nonumber & d_{3z^{2}-r^{2}}  = |20\rangle\\
\nonumber & d_{x^{2}-y^{2}}  =  \frac{1}{\sqrt{2}} (|22\rangle+|2\!-\!2\rangle)\\
\end{align}

Combining Eqs~\ref{C3states}, \ref{Ostates} and \ref{orbd}, we obtain the expression of the \textit{d}-functions, which are basis of the irreducible representations in $C_{3}$, in function of the \textit{d}-orbitals in $O$: 

\begin{eqnarray}
\nonumber &e_+(e) = {|2211\rangle}_3 = \frac{1}{\sqrt{2}} d_{3z^2-r^2} - \frac{i}{\sqrt{2}} d_{x^2-y^2}\\
\nonumber &e_-(e) = {|221\!-\!1\rangle}_3 = \frac{1}{\sqrt{2}} d_{3z^2-r^2} + \frac{i}{\sqrt{2}} d_{x^2-y^2}\\
\nonumber &a_1(t_2) = {|2\tilde{1}00\rangle}_3 = \frac{1}{\sqrt{3}} d_{xy} + \frac{1}{\sqrt{3}} d_{xz} -\frac{1}{\sqrt{3}} d_{yz}\\
\nonumber &e_+(t_2) = {|2\tilde{1}11\rangle}_3 =  \frac{1}{\sqrt{3}} d_{xy} + (-\frac{1}{2\sqrt{3}} + \frac{i}{2}) d_{xz} \\
& \nonumber ~~~~~~ + (\frac{1}{2\sqrt{3}} + \frac{i}{2}) d_{yz}\\
\nonumber &e_+(t_2) = {|2\tilde{1}1\!-\!1\rangle}_3 = \frac{1}{\sqrt{3}} d_{xy} - (\frac{1}{2\sqrt{3}} + \frac{i}{2}) d_{xz} \\
& ~~~~~~+ (\frac{1}{2\sqrt{3}} - \frac{i}{2}) d_{yz}
\label{states}
\end{eqnarray}

In $D_{3}$, ($e_+(t_2)$, $e_-(t_2)$) is basis of the irreducible representation $e(t_{2})$. Similarly, ($e_+(e)$ , $e_+(e)$) is a basis of the irreducible representation $e(e)$. In $D_{3}$, a mixing is thus possible between the functions belonging to the two $e$ irreducible representations, originating from the $e$ and $t_2$ levels in $O$.

\subsection{Definition of the crystal-field parameters used in the LFM calculations}

In $SO_{3}$ symmetry, the crystal-field Hamiltonian can be written as a combination of the \textit{q} components of unit tensors $U^{(k)}$ with rank \textit{k}.  Each tensor $U^{(k)}$ is associated to the \textit{k} irreducible representation of $SO_{3}$. For the Cr$^{3+}$ ion, \textit{k} = 0, 2 or 4, the term \textit{k} = 0 contributing only to the average energy of the configuration. Again, we will not make use of the parity ($\pm$ or \textit{g/u}). To study the Cr$^{3+}$ ion in trigonal symmetry $D_{3}$, we need to consider the subduction $SO_3\supset O \supset D_{3} \supset C_{3}$.

The crystal-field Hamiltonian is expressed as: 
$$H_{cc}
          = \sum_{k=2,4}X^{k(SO_3)\rho(O)\sigma(D_{3})\lambda(C_{3})}
             U^{k(SO_3)\rho(O)\sigma(D_{3})\lambda(C_3)}.$$
The unit tensor $U^{k(SO_3)\rho(O)\sigma(D_{3})\lambda(C_{3})}$ is related to the $\lambda$ irreducible representation of $C_{3}$ subgroup, coming from the \textit{k} irreducible representation of $SO_{3}$, that becomes $\rho$ in \textit{O}, $\sigma$ in $D_{3}$ and $\lambda$ in $C_{3}$. The terms $X^{k(SO_3)\rho(O)\sigma(D_{3})\lambda(C_{3})}$ are the crystal-field parameters used in the LFM code. Their definition, in function of ($D_{\sigma}$, $D_{\tau}$, $D_{q}$) or ($\nu$, $\nu$', $D_{q}$'), are given in Appendix G.\\
The branching rules which give 0 as irreducible representation in $D_{3}$ are summarized in Tab.~\ref{k=4}. 
\begin{table}[!]
\caption{Branching rules giving 0 as irreducible representation in $D_{3}$}
\label{k=4}
$$
\begin{array}{|c @{\quad \rightarrow \quad} 
               c @{\quad \rightarrow \quad}  c @{\quad \rightarrow \quad}
                              c @{\quad} |c|}
\hline
  SO_3 & O & D_{3}  & C_{3}
& X^{k(SO_3)\rho(O)\sigma(D_{3})\lambda(C_{3})}\\
\hline
  4 &    0      &     0     & 0 & X^{4000} \\
  4 & \tilde{1} &     0      & 0 & X^{4\tilde{1}00} \\ 
  2 & \tilde{1} &     0      & 0 & X^{2\tilde{1}00} \\ 
\hline
\end{array}
$$
\end{table}
This implies that : 
\begin{equation}
H_{cc} 
          = X^{4000} U^{4000} + X^{4\tilde{1}00}U^{4\tilde{1}00} + X^{2\tilde{1}00}U^{2\tilde{1}00}.
\end{equation}

Using the expression of the \textit{d}-eigenstates in $C_{3}$ given in Appendix E and the Wigner-Eckhart theorem (Eq. 4.2.4 of                                                                                                                          Ref.~\onlinecite{Butler}), we can now calculate the matrix elements 
\begin{align}
\nonumber &\langle k_1(SO_3)\rho_1(O)\sigma_1(D_{3})\lambda_1(C_{3})|H_{cc}|k_2(SO_3)\rho_2(O)\sigma_2(D_{3})\lambda_2(C_{3})\rangle\ .
\end{align}\\
We obtain the following equations: 
\begin{align}
\label{trig}
\nonumber \langle e_{\pm}(t_{2})|H_{cc}|e_{\pm}(t_{2})\rangle   = & -\frac{\sqrt{2}}{3\sqrt{15}}X^{4000}  + \frac{\sqrt{2}}{3\sqrt{21}}X^{4\tilde{1}00}\\
\nonumber & -\frac{1}{\sqrt{70}}X^{2\tilde{1}00}\\
\nonumber \langle a_{1}(t_{2})|H_{cc}|a_{1}(t_{2})\rangle  = & -\frac{\sqrt{2}}{3\sqrt{15}}X^{4000}  - 2\frac{\sqrt{2}}{3\sqrt{21}}X^{4\tilde{1}00} \\
\nonumber &+\frac{2}{\sqrt{70}}X^{2\tilde{1}00}\\
\nonumber \langle e_{\pm}(e)|H_{cc}|e_{\pm}(e)\rangle  = & \frac{1}{\sqrt{30}}X^{4000} \\
\nonumber \langle e_{\pm}(t_{2})|H_{cc}|e_{\pm}(e)\rangle  = & \frac{1}{2\sqrt{21}}X^{4\tilde{1}000} +\frac{1}{\sqrt{35}}X^{2\tilde{1}00}\\
\end{align}

\subsection{Relations between the crystal-field parameters used in LFM calculations and those derived from optical absorption spectroscopy}

As mentioned in Ref.~\onlinecite{Koenig}, two equivalent parameter sets are available in optical absorption spectrocopy to describe the crystal-field in trigonal symmetry: ($D_{\sigma}$, $D_{\tau}$, $D_{q}$) and ($\nu$, $\nu$', $D'_{q}$). In the following, we make connexion between the two sets.

\subsubsection {($D_{\sigma}$, $D_{\tau}$, $D_{q}$) parameter set}

$H_{cc}$ is defined as $H_{cc} = H_{cub} + H_{trig}$, where $H_{cub}$ and $H_{trig}$ are the Hamiltonian for the cubic and the trigonal contributions to the crystal-field, respectively. According to Ref.~\onlinecite{Koenig} (Eq. 3.88), we have:

\begin{align}
\nonumber &\langle e_{\pm}(t_{2})|H_{trig}|e_{\pm}(t_{2})\rangle  = D_{\sigma} + \frac{2}{3} D_{\tau}, \\
\nonumber &\langle a_{1}(t_{2})|H_{trig}|a_{1}(t_{2})\rangle  = -2D_{\sigma} -6D_{\tau}, \\
\nonumber &\langle e_{\pm}(e)|H_{trig}|e_{\pm}(e)\rangle  =  \frac{7}{3}D_{\tau}, \\
\nonumber &\langle e_{\pm}(t_{2})|H_{trig}|e_{\pm}(e)\rangle  =  -\frac{\sqrt{2}}{3}(3D_{\sigma} -5D_{\tau}). \\
\end{align}

If the cubic term $H_{cub}$ is added, we have: 

\begin{align}
\label{Koenig1}
\nonumber &\langle e_{\pm}(t_{2})|H_{cc}|e_{\pm}(t_{2})\rangle  = -4D_{q} + D_{\sigma} + \frac{2}{3} D_{\tau}, \\
\nonumber &\langle a_{1}(t_{2})|H_{cc}|a_{1}(t_{2})\rangle  = -4D_{q} - 2D_{\sigma} -6 D_{\tau}, \\
\nonumber &\langle e_{\pm}(e)|H_{cc}|e_{\pm}(e)\rangle  = 6D_{q}  + \frac{7}{3} D_{\tau},\\
\nonumber &\langle e_{\pm}(t_{2})|H_{cc}|e_{\pm}(e)\rangle  =   - \frac{\sqrt{2}}{3}(3D_{\sigma} -5D_{\tau}). \\
\end{align}

Combining Eqs~\ref{trig} and \ref{Koenig1}, we obtain: 

\begin{align}
\label{Dtau}
\nonumber &  X^{4000} = \frac{\sqrt{10}}{\sqrt{3}}(18D_{q} + 7D_{\tau}), \\
\nonumber & X^{4\tilde{1}00}  = \frac{10\sqrt{14}}{\sqrt{3}} D_{\tau}, \\
\nonumber & X^{2\tilde{1}00}  = -\sqrt{70} D_{\sigma}. \\
\end{align}

$X^{4000}$ contains only the cubic part of the crystal field, although $D_{\tau}$ appears in its expression. This means that the trigonal distortion, via $D_{\tau}$,  contributes to the $e-t_2$ splitting of the \textit{d}-orbitals. On the contrary, $X^{4\tilde{1}00}$ and $X^{2\tilde{1}00}$ are entirely due to the trigonal distortion. 
Hence, when we investigated the effect of the trigonal distortion on the XANES spectra in the LFM calculations, the value of $X^{4\tilde{1}00}$ and $X^{2\tilde{1}00}$ were set to zero, while the value of $X^{4000}$ was fixed to the value used in $D_{3d}$ symmetry. 
Things can be simplified by defining $D'_q$, so that it contains also the contribution of $D_{\tau}$ to the cubic field. This leads to the definition of two other distortion parameters, $\nu$ and $\nu'$.

\subsubsection {($\nu$, $\nu'$, $D'_{q}$) parameter set}

The crystal-field Hamiltoninan $H_{cc}$ is now defined as $H_{cc} = H'_{cub} + H'_{trig}$, where $H'_{cub}$ contains the contribution of  $D_{\tau}$. According to Ref.~\onlinecite{Koenig} (Eq. 3.90), we have:

\begin{align}
\nonumber &\langle e_{\pm}(t_{2})|H'_{trig}|e_{\pm}(t_{2})\rangle  = - \frac{1}{3} \nu, \\
\nonumber &\langle a_{1}(t_{2})|H'_{trig}|a_{1}(t_{2})\rangle  = \frac{2}{3} \nu, \\
\nonumber &\langle e_{\pm}(t_{2})|H'_{trig}|e_{\pm}(e)\rangle  =  \nu' .\\
\end{align}

If we add the cubic term $H'_{cub}$, which is here defined so that the center of gravity is maintained for the trigonally split $t_{2g}$ orbitals, the following equations are obtained: 

\begin{align}
\label{Koenig2}
\nonumber &\langle e_{\pm}(t_{2})|H_{cc}|e_{\pm}(t_{2})\rangle  = -4D'_{q} - \frac{1}{3} \nu,\\
\nonumber &\langle a_{1}(t_{2}|H_{cc}|a_{1}(t_{2})\rangle  = -4D'_{q} + \frac{2}{3} \nu, \\
\nonumber &\langle e_{\pm}(t_{2})|H_{cc}|e_{\pm}(e)\rangle  =   \nu'. \\
\end{align}

Combining Eqs \ref{trig} and \ref{Koenig2}, we have: 

\begin{align}
\label{nu}
\nonumber &  X^{4000} = 6 \sqrt{30} D'_{q}, \\
\nonumber & X^{4\tilde{1}00}  = -\frac{2\sqrt{3}}{\sqrt{7}} (\sqrt{2}\nu-3\nu'), \\
\nonumber & X^{2\tilde{1}00}  = \frac{4\sqrt{35}}{7}(\nu'+\frac{1}{2\sqrt{2}}\nu). \\
\end{align}

\subsubsection {Relations between the two parameter sets}

Combining Eqs \ref{Dtau} and \ref{nu}, we obtain the relations given in Ref.~\onlinecite{Koenig} (Eq.~3.91):

\begin{align}
\nonumber &  D'_{q} = D_{q} +\frac{7}{18} D_{\tau},\\
\nonumber &  \nu  = -3D_{\sigma} - \frac{20}{3} D_{\tau},  \\
\nonumber & \nu'  = -\sqrt{2}D_{\sigma} + 5\frac{\sqrt{2}}{3} D_{\tau}. \\
\end{align}

\newpage


\begin{thebibliography}{2}
\bibitem[14]{octa} Note that the direction of the chains has been incorrectly reported in Ref.~\onlinecite{Juhin}.
\end{thebibliography}
\end{document}